\documentclass[]{aa} \usepackage{graphicx}\usepackage{rotating}

\def\simlt{$\; \buildrel < \over \sim \;$}
\def\ltsima{\lower.5ex\hbox{\simlt}}
\def\simgt{$\; \buildrel > \over \sim \;$}
\def\deg{\degr}
\def\gtsima{\lower.5ex\hbox{\simgt}}

\def\etal{et~al.~}

\def\xmm{{\it XMM-Newton~}}
\def\integral{{\it INTEGRAL~}}
\def\sax{{\it BeppoSAX~}}
\def\chandra{{\it Chandra~}}
\def\rxte{{\it RXTE~}}
\def\asca{{\it ASCA~}}

\def\cm2{\rm{cm}^{2}}
\def\cmM2{\rm cm$^{-2}$}

\def\Ms{M_\odot}

\def\rerg{\rm erg}

\def\rcm{\rm cm}

\def\flux{\rm erg\ cm^{-2}\ s^{-1}}
\def\pflux{\rm  ph\ cm$^{-2}$\ s$^{-1}$}
\def\csq{\chi^2}

\def\apj{{ApJ}}
\def\apjs{{ApJS}}

\def\aa{{A\&A}}
\def\aas{{A\&AS}}
\def\mnras{{MNRAS}}

\def\41{NGC~4151}

\begin{document}

\title{ The broad band spectrum and variability of NGC 4151 observed by
BeppoSAX}

\author{  A. De Rosa  \inst{1} \and L. Piro \inst{1} \and G. C. Perola \inst{2}
\and M. Capalbi \inst{3} \and M. Cappi \inst{4} \and P. Grandi
\inst{4} \and L. Maraschi \inst{5} \and P. O. Petrucci \inst{6} }
\institute{ {Istituto di Astrofisica Spaziale e Fisica Cosmica, INAF, 
sezione di Roma, Via Fosso del Cavaliere, 00133 Roma, Italy} 
\and {Dipartimento di Fisica, Universit\`a degli
Studi ``Roma Tre'', Via della Vasca Navale 84, I--00146 Roma,
Italy}
\and
{ASI Science Data Center, c/o ESA-ESRIN, Via Galileo Galilei, 00044 
Frascati, Italy}
\and
{Istituto di Astrofisica Spaziale e Fisica Cosmica, INAF, sezione di 
Bologna, Via Gobetti 101, I-40129,Italy}
\and
{Osservatorio Astronomico di Brera, INAF, Via Brera 28, Milan I-20121, Italy}
\and{Laboratoire d'Astrophysique de Grenoble, BP 43, 38041 
Grenoble Cedex 9, France }}

\offprints{Alessandra De Rosa: alessandra.derosa@iasf-roma.inaf.it}
\date{ }
\titlerunning{BeppoSAX observations of NGC 4151}
\maketitle

\begin{abstract}

We present an analysis of all \sax observations of \41.
This source was observed 5 times from 1996 to 2001 with durations
ranging from a day to four days. The intrinsic continuum
(described as a cut-off power law), is absorbed at low energies
by a complex system: a cold patchy absorber plus a warm uniform screen 
photoionized by the central continuum. We find that this ``dual absorber'' is
the main driver of the observed variability, up to a factor
of eight, at 3  keV. In particular the covering fraction of the
cold absorber changes on time
scales of the order of a day, supporting its association with the
Broad Line Region. The column
density of the warm gas varies on a longer time scale (months to year). 
Some of the small amplitude spectral variability above 10 keV can be
explained with an intrinsic variation (with variation of the photon index $\Delta\Gamma \sim
0.2$). The flux below 1 keV remains constant confirming an extended origin.
Its spectrum is reproduced by a combination of a thermal component 
(with temperature $kT=0.15$ keV)
and a power law with the same slope as the
intrinsic continuum but with an intensity a few per
cent. A Compton reflection component is significantly detected
in 1996 (averaged value of $\Omega/2\pi \sim0.4$, with $\Omega$ the
solid angle covered by the reflecting medium), with intensity decreasing on
time scale of year, and it desappears in 2000 and 2001.
The long time scale of variations argues for an association with an 
optically thick torus at a distance of few light years. 
An iron line was detected in all spectra. Its energy is
consistent with fluorescence by cold iron. 
We find that the line is variable. Its behaviour is reproduced by a variable component
proportional to the level of the reflection flux plus a constant component. The
flux of the latter is consistent with the extended line emission
observed by {\it Chandra}.
We conclude that the first component is likely arising from the torus and the
second is produced in the extended Narrow Line Region.

\end{abstract}


\section{Introduction}

The Seyfert galaxy \41 is one of the brightest AGN. It has been
considered for years the prototype of its class and, as such, has
been extensively studied at all wavelengths (see Ulrich 2000 for
a review). In X-rays its spectrum  is the most complex observed so
far in AGN, being characterized by narrow and broad spectral
features from soft to hard X-rays (e.g. \cite{per86}; \cite{zdz96}; \cite{zdz02}).
The central X-ray continuum is absorbed below a few keV by a
(complex) absorber with column density $N_{\rm H}\approx10^{23}$
\cmM2 associated with the Broad Line Region (BLR,
\cite{reichert}). This absorption in \41
 allows to reveal the presence of soft X-ray components
below 1 keV, associated with a scattering component and a thermal plasma
with a low temperature of $\sim 0.1$ keV (\cite{per86};
\cite{pou86}; \cite{wb}) as those observed in  type 2 objects
(e.g. \cite{antonucci}; \cite{matt_s2}).

At higher energies \41 shows an iron line which remains remarkably
constant notwithstanding large variations of the continuum
(\cite{per86}; \cite{schurch03}). Observations by \chandra
(\cite{chandra}) have indeed shown that most of the line is
produced in the Narrow Line Region (NLR). The iron line site of
Seyfert 1 galaxies has often been associated in the past with an
accretion disk (e.g. \cite{tanaka}; \cite{fabian2000}). 
\xmm and \chandra observations have shown that a narrow line
component is a common feature in Sy 1s (Reeves \etal 2001; Pounds
\etal 2001; Matt \etal 2001; Kaspi \etal 2001). 
Further insight on the nature of the
reprocessing medium should also become available through
measurements of the reflection component.
Recent \integral observation (\cite{beckmann2005}) described the hard X-ray/soft-gamma
spectrum of \41 with a Compton continuum from hot electrons 
($kT_e\simeq $100
keV) in an optically thick ($\tau$=1.3) corona plus a reflected
component from a cold material subtending a solid angle of $\Omega/2\pi
\simeq 0.7$. 

NGC 4151 has been
one of the first Seyfert 1 galaxy with a well established evidence of changes
of the spectral index correlated with the luminosity
(\cite{per86}; \cite{yaq93}).
With the
increased bandwidth and sensitivity of present instrumentation,
this feature appears to be present in  other Seyfert 1 galaxies
(e.g. 1H~0419-577, Page \etal 2002; NGC~3783, De Rosa \etal 2002;
MCG-6-30-15, Vaughan \& Edelson 2001;
NGC~5548, Petrucci \etal 2000, Nicastro \etal 2000; NGC~7469,
Nandra \etal 2000; IC~4329A, Done \etal 2000; Mkn~509, De Rosa \etal
2004, \cite{zdz03}).
In summary, in the broad X-ray range from 0.1 to 200 keV, \41
represents a case study of the common properties
(absorption/scattering  medium) of type 1 and
type 2 objects, of AGN environment from light-day to kpc,
and of investigation of the properties of the intrinsic continuum.

In this paper we present the results  of the complete set of
observations by \sax performed in 1996, 1999, 2000 and 2001.
The observations and data reduction are described in Section
\ref{observations}.
A model-independent variability study is presented in Section \ref{specratios}.
The broad-band spectral analysis is reported in Section
\ref{bb spectral model} and discussed in Section \ref{discussion}.
Our conclusions are drawn in Section \ref{conclusions}.

\section{\sax observations and data reduction}
\label{observations}

\41 was observed by \sax (\cite{psb}; \cite{boella}) several
times: in  July 1996, December 1996, January 1999, December 2000 and
December 2001. The log of the observations is given in Table \ref{log}.

The first observation, part of the Science Verification Phase (SVP), 
was originally planned to
last for about 3 days. However the
pointing was interrupted for 1 day and then restarted 2 days after. In
the  observations in 1996 the three MECS units, the HPGSPC and the
PDS were operated. The LECS was always on, with the exception of
the second part of the observation in July 1996. From January 1999
onward two MECS units, the LECS and the PDS were operating.

\begin{figure*}
\begin{center}
\includegraphics[height=6.cm,width=6.cm]{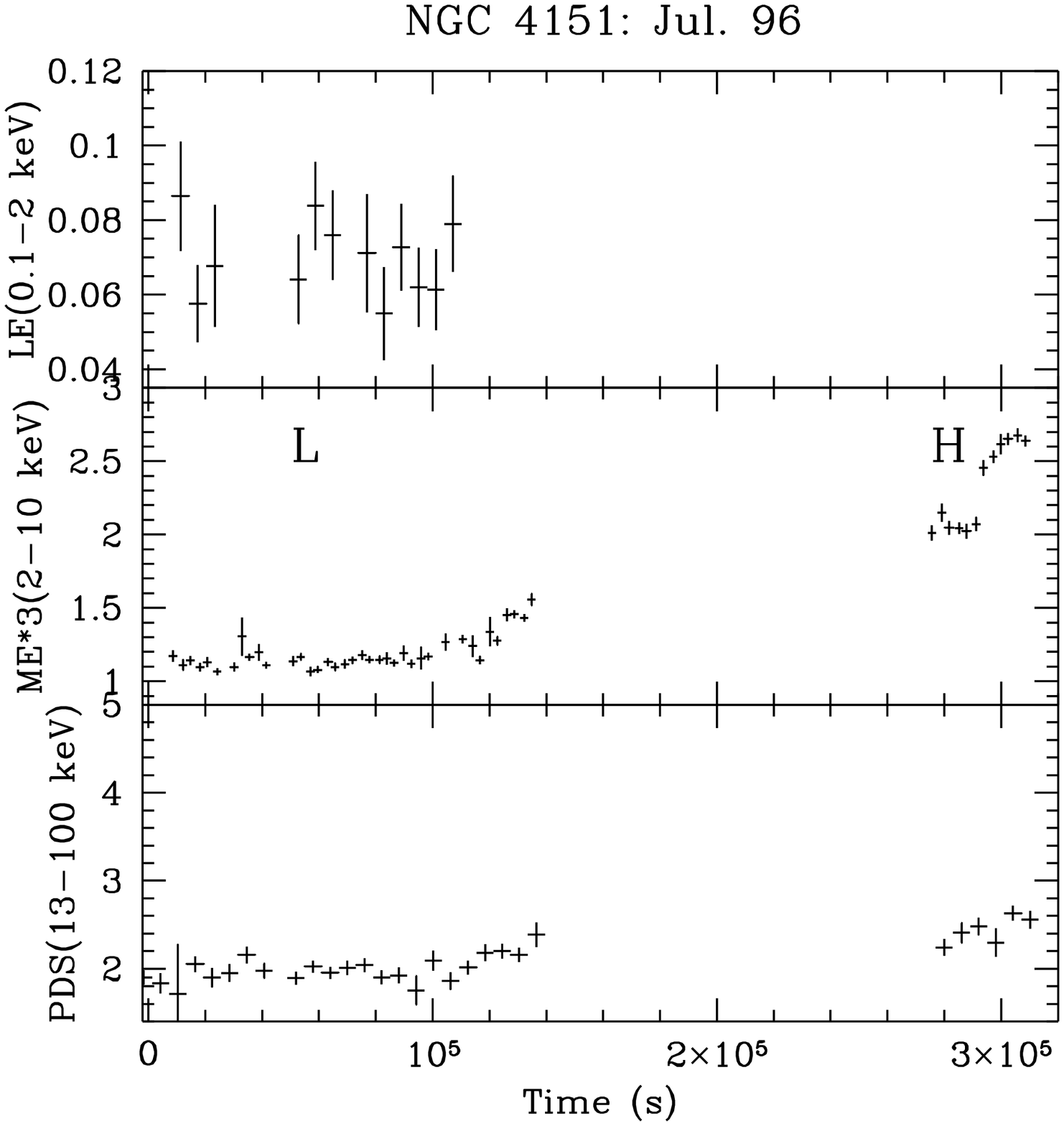}
\includegraphics[height=6.cm,width=6.cm]{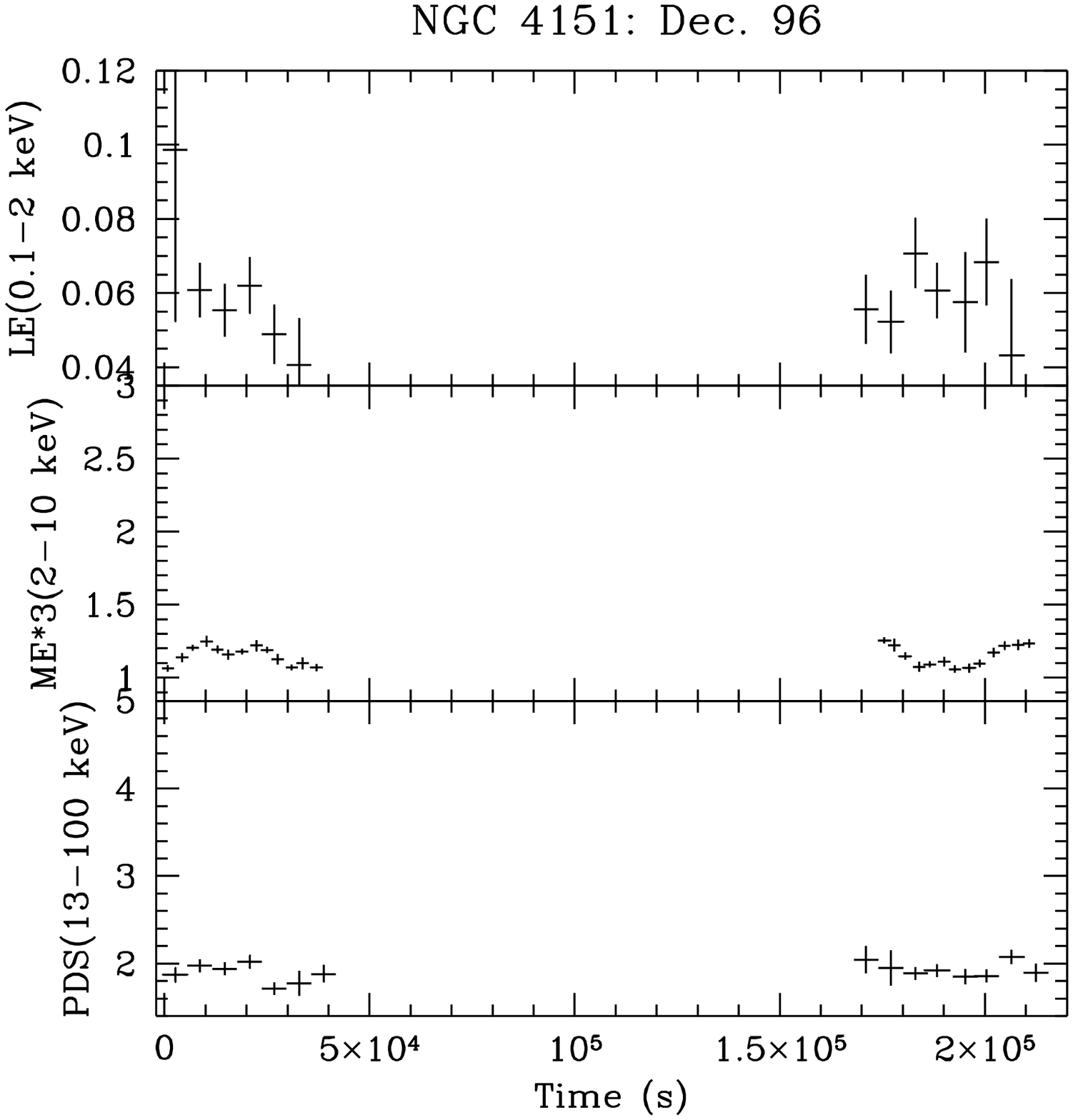}
\end{center}
\begin{minipage}{20cm}
\includegraphics[height=6.cm,width=5.5cm]{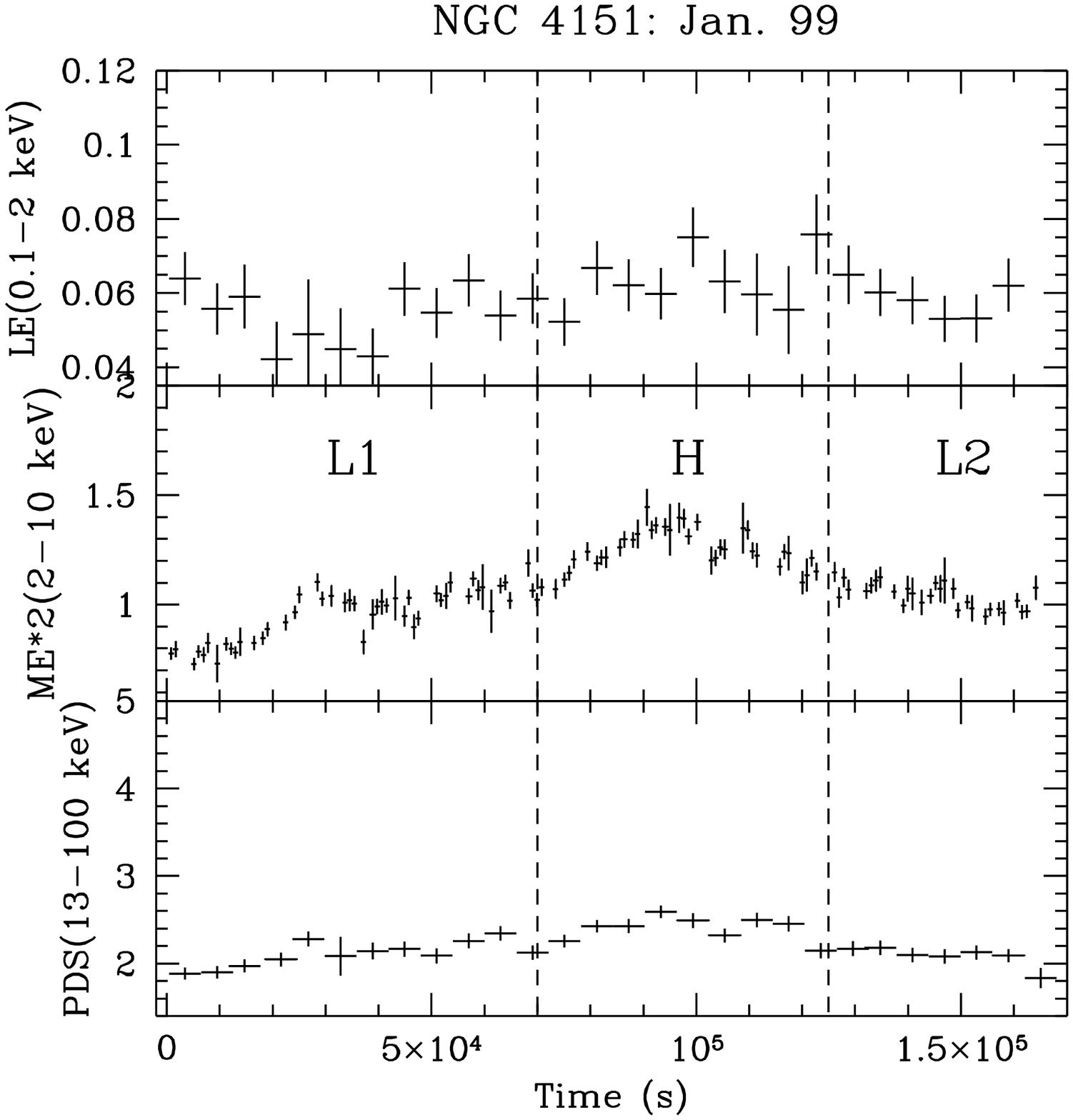}
\includegraphics[height=6.cm,width=5.5cm]{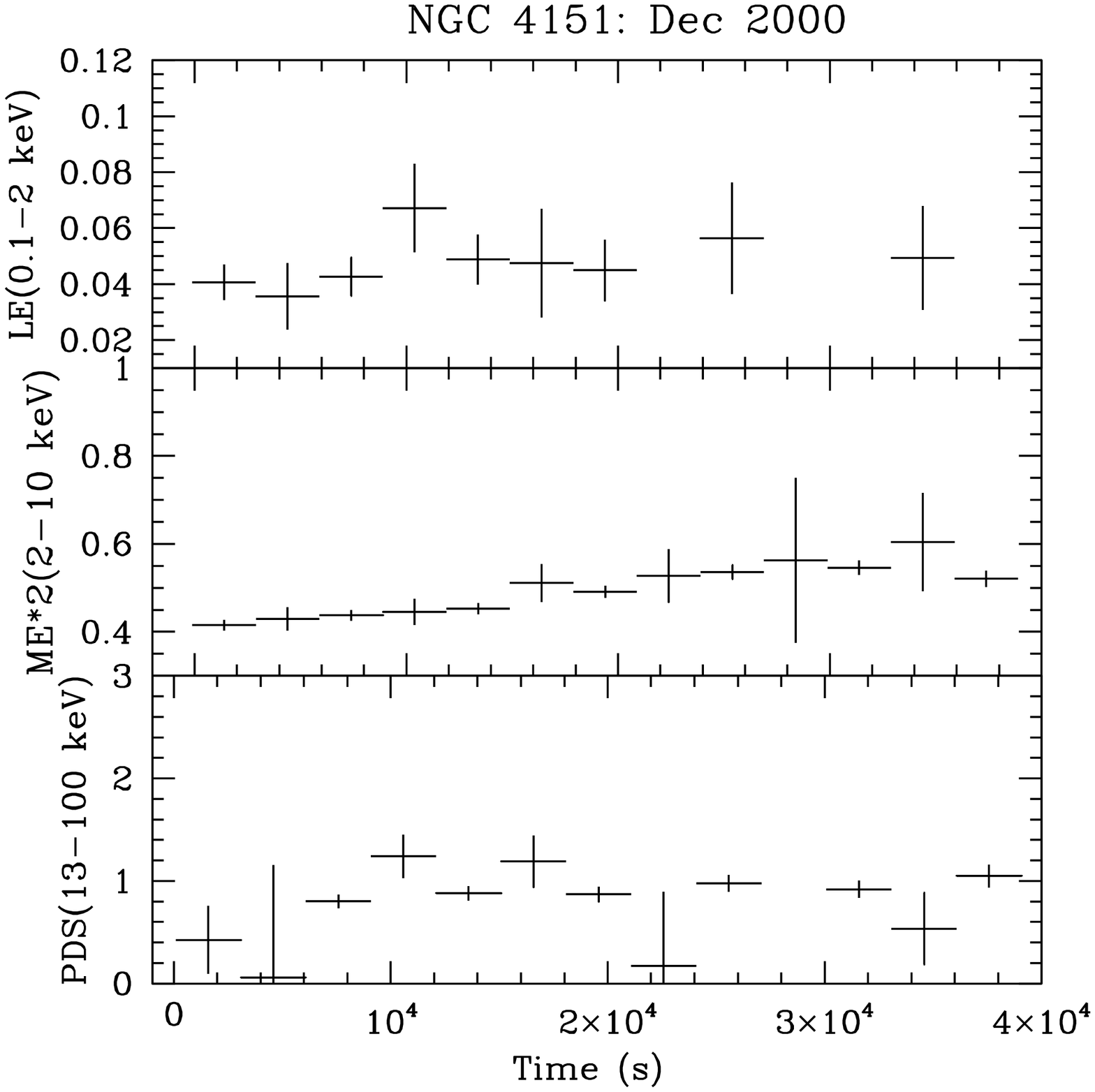}
\includegraphics[height=6.cm,width=5.5cm]{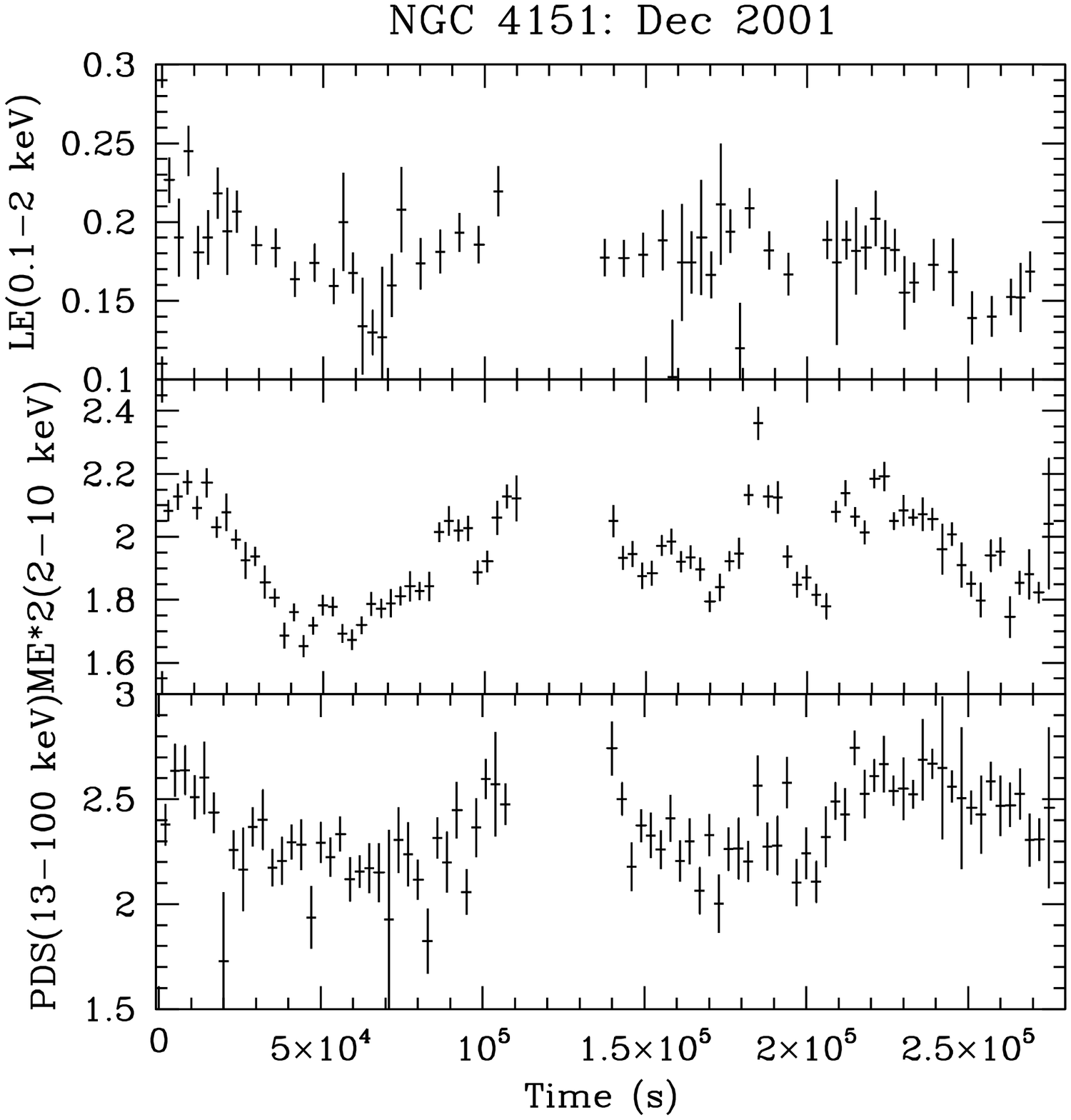}
\end{minipage}
\caption[]{Light curve of NGC4151 in cts/s. We note that from 1999
  onward (lower panels)
two MECS units were operating, while in 1996 (upper panels) they were three.}
\label{lc}
\end{figure*}

The data of all instruments were reduced using  standard
procedures (\cite{fgg99}). A BL Lac object is located $\sim
5'$ away from NGC 4151. The count rates are  $\sim$ 2 per cent and $\sim$
20 per cent of that of \41 for the MECS and LECS respectively. For
the MECS we adopted the standard extraction circle of $4'$, in which
the contribution of the BL Lac  is negligible ($\ltsima 0.4$ per cent).
In the extraction of spectra and light curves of LECS we excluded
a circle of $2'$ of radius centered on the BL Lac object, while
adopting the standard source extraction radius of $8'$. We estimate
a residual contribution of the BL Lac of $\approx 5-10$ per cent,
comparable to the statistical errors. The PDS spectra filtered
either with fixed or variable rise time thresholds did not show
any significant difference. The latter method is more suited to
sources fainter than \41, so we have adopted data derived
with the first method.\\
In Figure \ref{lc} we show the  light curves of the LECS (0.1--2
keV), MECS (2--10 keV) and PDS (13--100 keV). The MECS (2--10 keV)
counts rate during the July 1996 observation changed  by a factor of
two on time scales of one day,  from a low state (hereafter J96L)
to a high state (J96H). The intensity during the December 1996
observation was constant (D96), and quite similar
to the J96L state, then low. Also the observation in December 2000 (D00)
 was constant. In January 1999 the average flux was at an
intermediate level between the Low and High states defined above.
During the observation the source flux exhibited a slow increase
in the first part of the observation (J99L1), reached a peak
(J99H) and then started to decrease (J99L2). The time scale  was
$\sim 0.5\ $ days, with an amplitude of $\sim 1.8$ (max to min).
The PDS (13--100 keV) light curve follows the same pattern of
variability of the MECS, but with a reduced amplitude: for example
the ratio  high to low states in July 1996 is $2.00\pm0.01$ for the
MECS and $1.22\pm0.03$ for the PDS (Table \ref{log}). On the
reverse the LECS (0.1--2 keV) count rate, with the exception of December 
2001 (D01), does not show significant variability: the small difference 
($<20$ per cent, see Table \ref{log}) of level between different observations     
is consistent with the
effect of grid obscuration of the source (\cite{fgg99}), placed in
different positions in the three observations. The different behaviour in 
December 2001 will be discussed in Section \ref{sec:softdata}.
Considering the substantial amount of spectral variability shown
by this object, we have produced  separate spectra for each of the
above mentioned states, in which (2--10 keV) variations are
$\ltsima 25$ per cent.\\
Hereafter errors and upper limits on spectral parameters
correspond to  $\Delta \csq=2.7$, i.e. 90 per cent confidence level for a
single parameter of interest. Spectral power indexes refer to
photon distribution. In spectral fitting the normalizations of
LECS, HPGSPC and PDS relative to the MECS have been left free to
vary within the current ranges of uncertainty: LECS/MECS: 0.7--1.0;
HPGSPC/MECS=0.95--1.1; PDS/MECS=0.77--0.95.

\begin{sidewaystable*}
\centering
\caption[] {Log of the observations}
\begin{tabular}{l|cc|cc|cc|cc}
\noalign {\medskip}
\hline
 {\bf Date}&  & \bf LECS (0.1--2 keV) & & \bf MECS (2--10 keV) &
 &\bf HPGSPC (8--20 keV) & & \bf PDS (13--200 keV) \\
 & $t_{\rm exp}^1$ & cnt rate$^2$ &  $t_{\rm exp}^1$ & cnt rate$^2$  &  $t_{\rm exp}^1$ & cnt
 rate$^2$  &  $t_{\rm exp}^1$ & cnt rate$^5$ \\ \hline
6-10/7/96 (J96L) & 6 & $0.065\pm0.003$ & 56 & $1.145\pm0.005^3$ & 26 &
$1.86\pm0.06$ & 26 & $2.36\pm0.02$ \\ 
~~~``~~~~ (J96H) &- &- & 15 & $2.29\pm0.01^3$ &7 &$2.7\pm0.1$ & 8 & $2.90\pm0.04$\\
4-7/12/96 (D96)& 13 & $0.051\pm0.003$ & 48 & $1.102\pm0.005^3$ & 23 & $1.9\pm0.06$ & 19 & $2.28\pm0.03$\\
4-6/1/99 (J99L1) & 14 & $0.053\pm0.002$ & 36 & $0.937\pm0.005^4$&- & -& 16 & $2.55\pm0.04$\\
~~~``~~~~ (J99H) & 10 & $0.057\pm0.003$ & 27 & $1.238\pm0.007^4$ &- &- & 13 & $2.95\pm0.04$\\
~~~``~~~~ (J99L2) & 10 &$0.059\pm0.002$ & 20 & $1.015\pm0.007^4$ &- &- &9 &$2.60\pm0.04$ \\
22-23/12/2000 (D00) & 4 & $0.035\pm0.003$ & 19 & $0.470\pm0.005^4$ & - & - & 9 & $1.01\pm0.04$ \\
18-21/12/2001 (D01) & 43 & $0.145\pm0.002$ & 114 & $1.929\pm0.004^4$ & - & - & 53 & $2.90\pm0.01$ \\
\noalign {\hrule}
\noalign{\medskip}
\noalign{\noindent
Note: $^1$ net exp. time in ks;
$^2$ s$^{-1}$. For the MECS we indicate the total rate of all MECS
unit operating;
$^3$ 3 MECS units;
$^4$ 2 MECS units;
$^5$  s$^{-1}$ per half detector}
\end{tabular}
\label{log}
\end{sidewaystable*}

\section{Spectral ratios}
\label{specratios} 

\begin{figure*}[] 
\vspace{-0.5cm}
\centering
\includegraphics[height=8.5cm,width=8.cm]{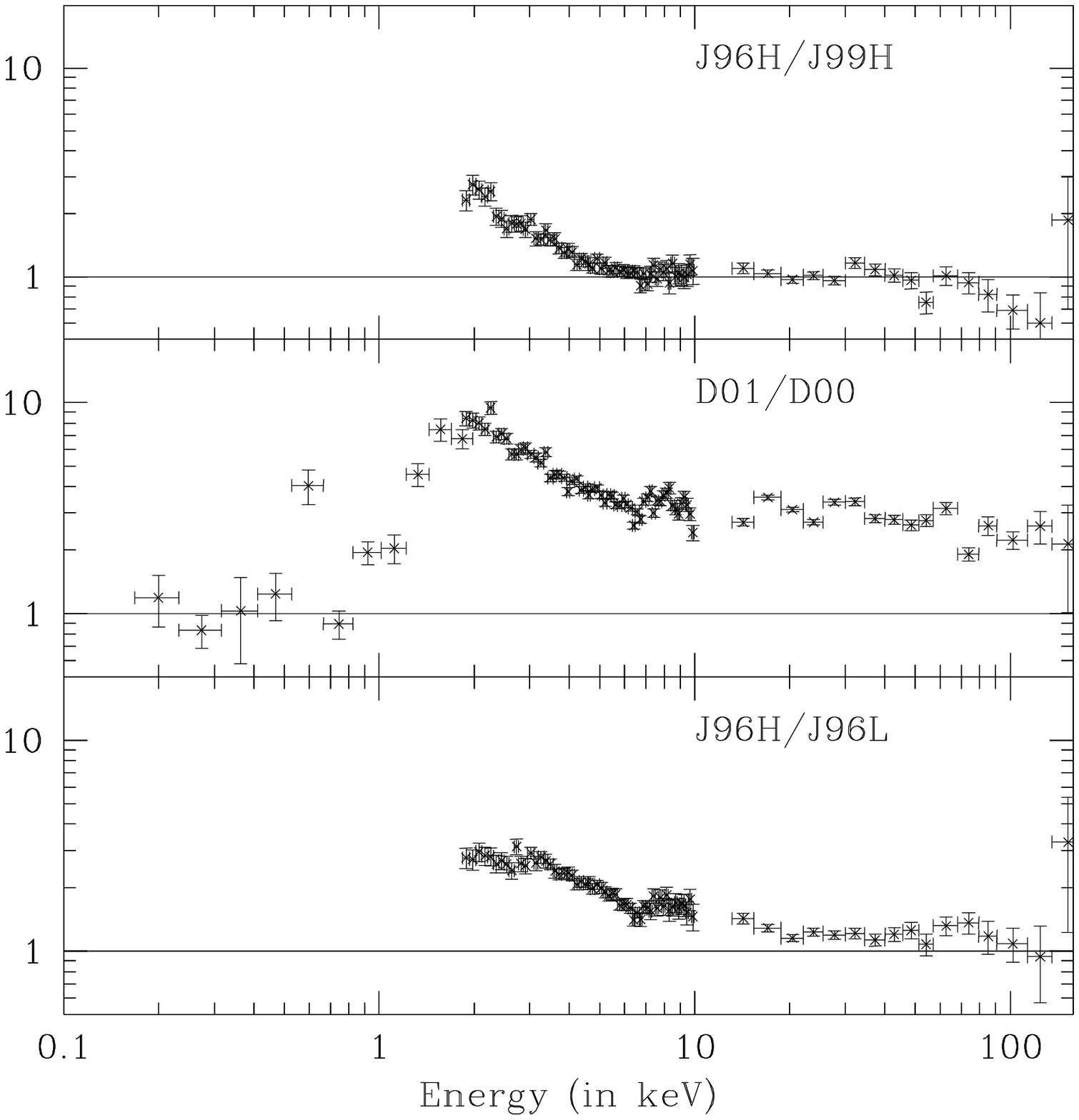}
\includegraphics[height=8.5cm,width=8.cm]{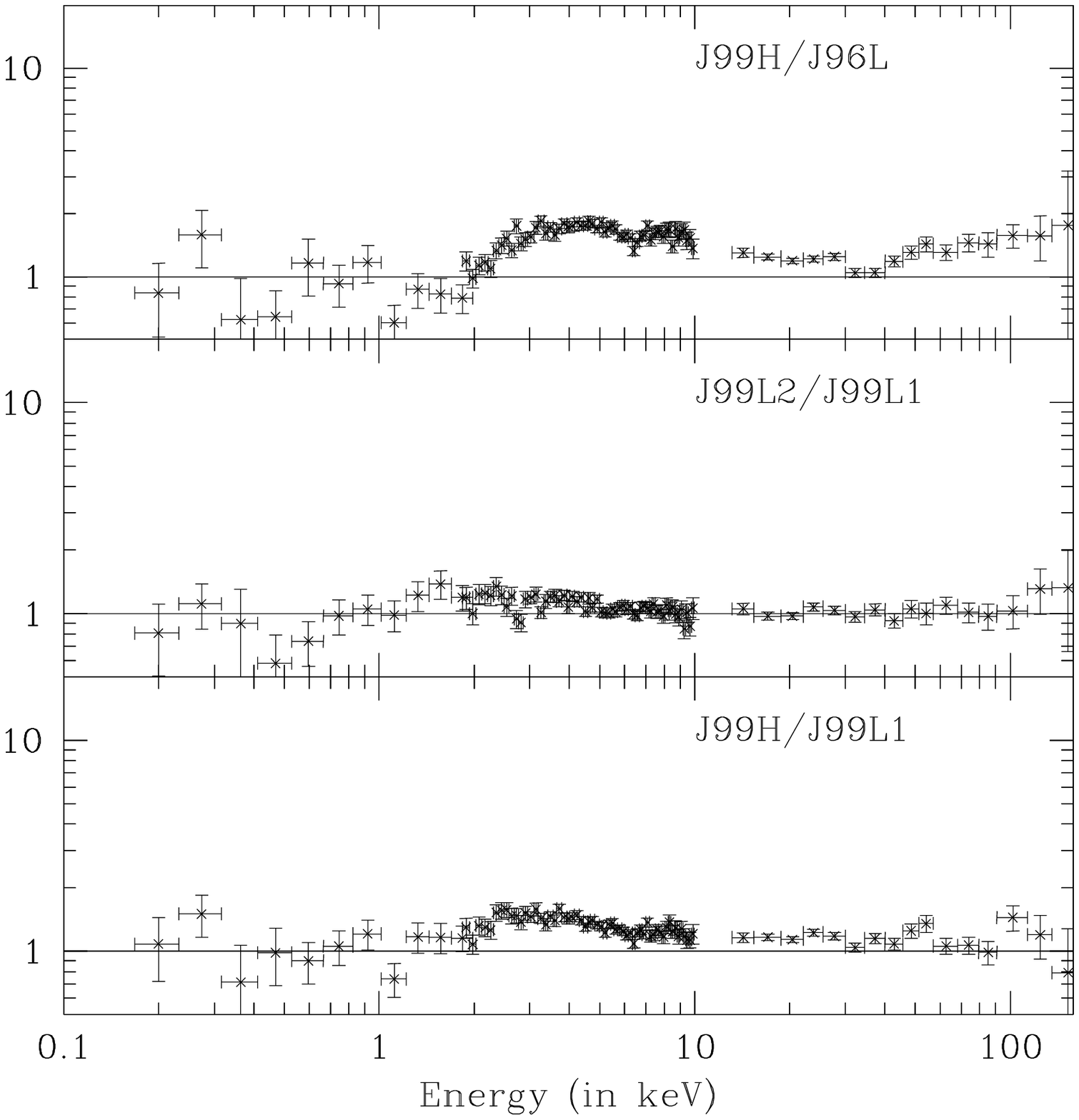}
 \caption{ Spectral ratio between  different states.}
\label{ratios}
\end{figure*}

The spectral ratios derived comparing the
spectral states observed during our campaign are presented in
Figure \ref{ratios}. This allows us to derive a model-independent
description on the spectral variability of the source. We note the
following:

\begin{enumerate}

\item Below 1 keV the flux remains constant, notwithstanding the
large variations showed at E$\gtsima$ 2 keV. The low energy
 spectrum is thus to be dominated by constant component(s)
(\cite{per86}; \cite{pou86}; \cite{zdz96}). This means that some absorber has to
suppress any variable intrinsic component  at low energies. 
Around 0.6--0.7 keV the ratio D01/D00 shows evidence of a changing of 
the opacities of the warm gas. We will discuss this effect in Section 
 \ref{sect:absdata}.

\item The 2--10 keV range is characterized by the largest
variability, up to a factor of $\sim 8$ in the 2--3 keV range. The
spectrum gets steeper when the flux increases (J96H/J96L,
J99H/J99L1, D01/D00, J99H/J96L). This happens both on short (days)
and long time scales (years). This behaviour has been attributed
to
 intrinsic spectral variations correlated with the luminosity
 on time scale as short as few hours
(\cite{yaq93}; \cite{per86}), {\it and} to variations of the absorber on
month-year time scales. Warwick \etal (1995) and Weaver \etal (1994b) instead attributed all
variations  to changes of absorber structure. We will assess this
issue through a broad band spectral fitting (see Section
\ref{spectral variability}).


\item
Above 10 keV two kind of behaviour are apparent. a) The spectral
ratios relative to observations performed within $\approx$ day are
all consistent with a constant (J96H/J96L: $\chi^2_\nu=1.2$,
J99L2/J99L1: $\chi^2_\nu=0.6$ and J99H/J99L1:
$\chi^2_\nu=1.2$). b) The spectral ratios relative to
observations performed over time scales of months-years are
not (J96H/J99H: $\chi^2_\nu=2.0$ , and D01/D00:
$\chi^2_\nu=1.8$,  and J99H/J96L: $\chi^2_\nu=2.6$).
This is a new result that, as it will be shown in Section \ref{reflection}, bears
important implications on the origin of the reflection component.

\end{enumerate}

In the next sections we will present a detailed quantitative
analysis of these results, through detailed spectral deconvolution.

\section{Broad-band (0.1--200 keV) spectral modelling}
\label{bb spectral model}

On the basis of previous
observations and models developped in previous published work, to fit the \sax broad-band spectra (with all
the instruments in the energy range specified in Table \ref{log}), we
adopted a spectral template model ({\it baseline model} BLM),  which includes the
following  components:

A. The intrinsic continuum is described by a power-law
with an e-folding energy $E_{\rm c}$. 

B. A Compton reflection component (PEXRAV model in XSPEC,
\cite{pexrav_ref}), with the cosine of inclination angle of the
reflector fixed to $0.95$. We assume elemental
abundances from Anders \& Grevesse (1989).

C. A narrow $K\alpha$ iron line modeled with a gaussian profile with
the intrinsic width set to 10 eV.

D. The soft X-ray spectrum ($E < 2$ keV) is a combination
of a thermal component model (MEKAL) with the temperature fixed to 0.15
keV and a scattering component (i.e.
a power-law with the same slope of the intrinsic hard continuum but
with a different normalization).

E. The complex absorber is reproduced with a dual absorber model. A fraction
$f_{\rm cov}$ of the source is covered by a cold column density $N_{\rm Hcov}$.
An additional uniform and photoionized gas is responsible for
additional absorption of the spectrum at low energies and for a midly
ionized Fe edge at 7.3--7.5 keV (FeVII-FeVIII) 
(ABSORI model in XSPEC \cite{done92}).

This photoionized gas is characterized by a temperature $T$=$3\times
10^4$ K, a column density $N_{\rm Hwarm}$
and a ionization parameter $\xi=L_{\rm ion}/nR^2$, where $L_{\rm ion}$ is
the source luminosity between 5 eV and 300 keV,  $n$ is the hydrogen
number density of the gas and $R$ its distance
from the central ionizing source.  The Fe abundances in the
cold and warm absorbers and in the reflection component are tied together
and left free during the fit.

Our model assumes that only the e-folded power-law
with reflection (i.e. the intrinsic continuum), is subject to
complex absorption. Additional absorption through the Galaxy is applied to
{\it all} the emission components ($N_{\rm Hgal}=2\times 10^{20}$ \cmM2).

\begin{figure*}[]
\centering
\begin{minipage}{15cm}
\includegraphics[height=11.cm,width=11.cm,angle=-90]{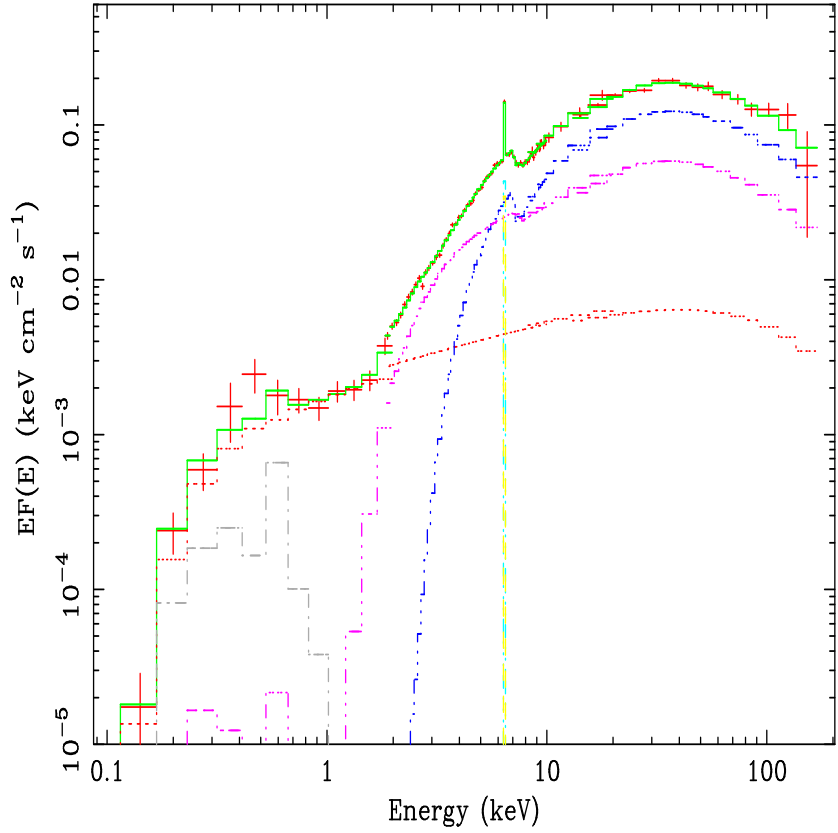}
\end{minipage}\vspace{1cm}
\begin{minipage}{15cm}
\includegraphics[height=11.cm,width=11.cm,angle=-90]{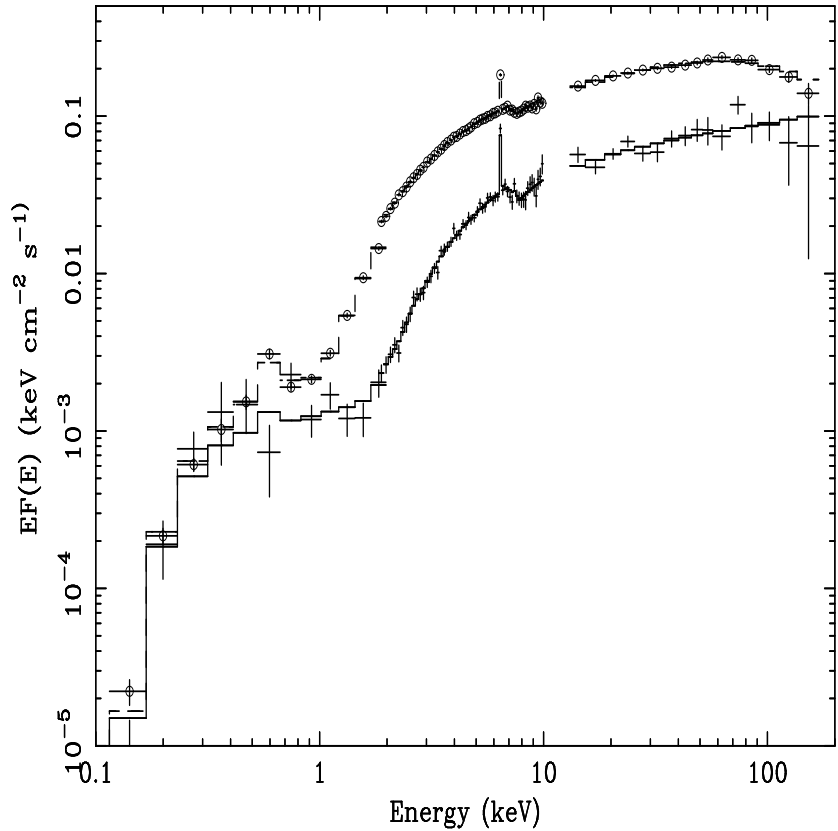}
\caption{\sax broad-band spectrum $E F(E)$. In the upper panel we plot the
  spectrum in July 1996 low state with all the spectral components of
  the BLM described in the text: in green (solid line) the total
  spectrum, in red (dot line) the scattered component, in grey
  (dash-dot line) the thermal soft component, in blue and magenta
  (dash-dot-dot lines) the transmitted components through
  the cold and photoionized gas respectively, 
  in yellow (dash-line) the iron line. In the lower panel we plot
  both spectra in December 2001 (dots) and December 2000 (crosses). 
  There for clarity we do not plot the different spectral components 
  but the total model only.}
\label{bbspec}
\end{minipage}
\end{figure*}

\begin{figure*}[!] 
\vspace{-0.5cm}
\centering
\includegraphics[height=9.5cm,width=9.cm]{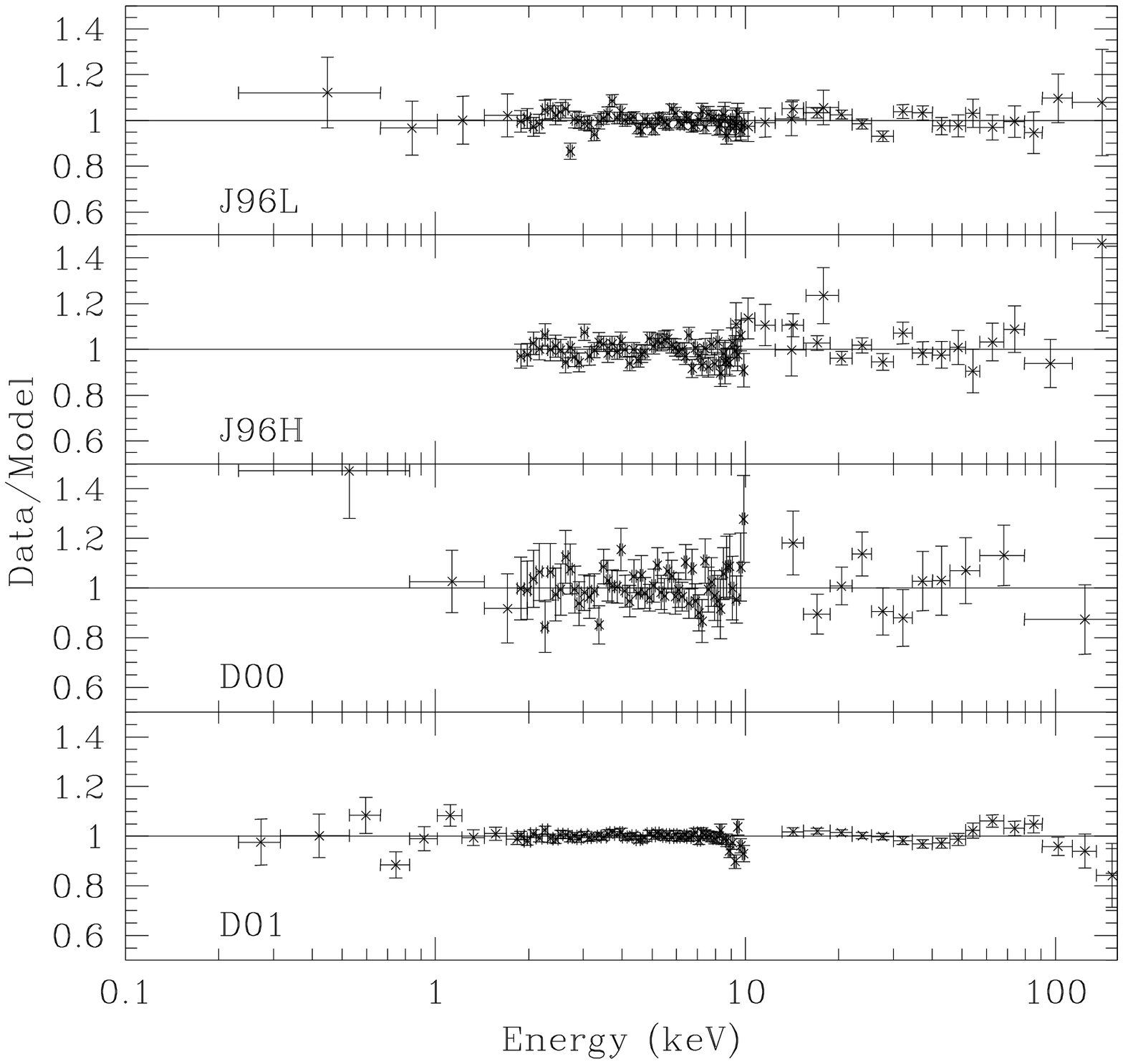}
\vspace{3cm}
\includegraphics[height=9.5cm,width=9.cm]{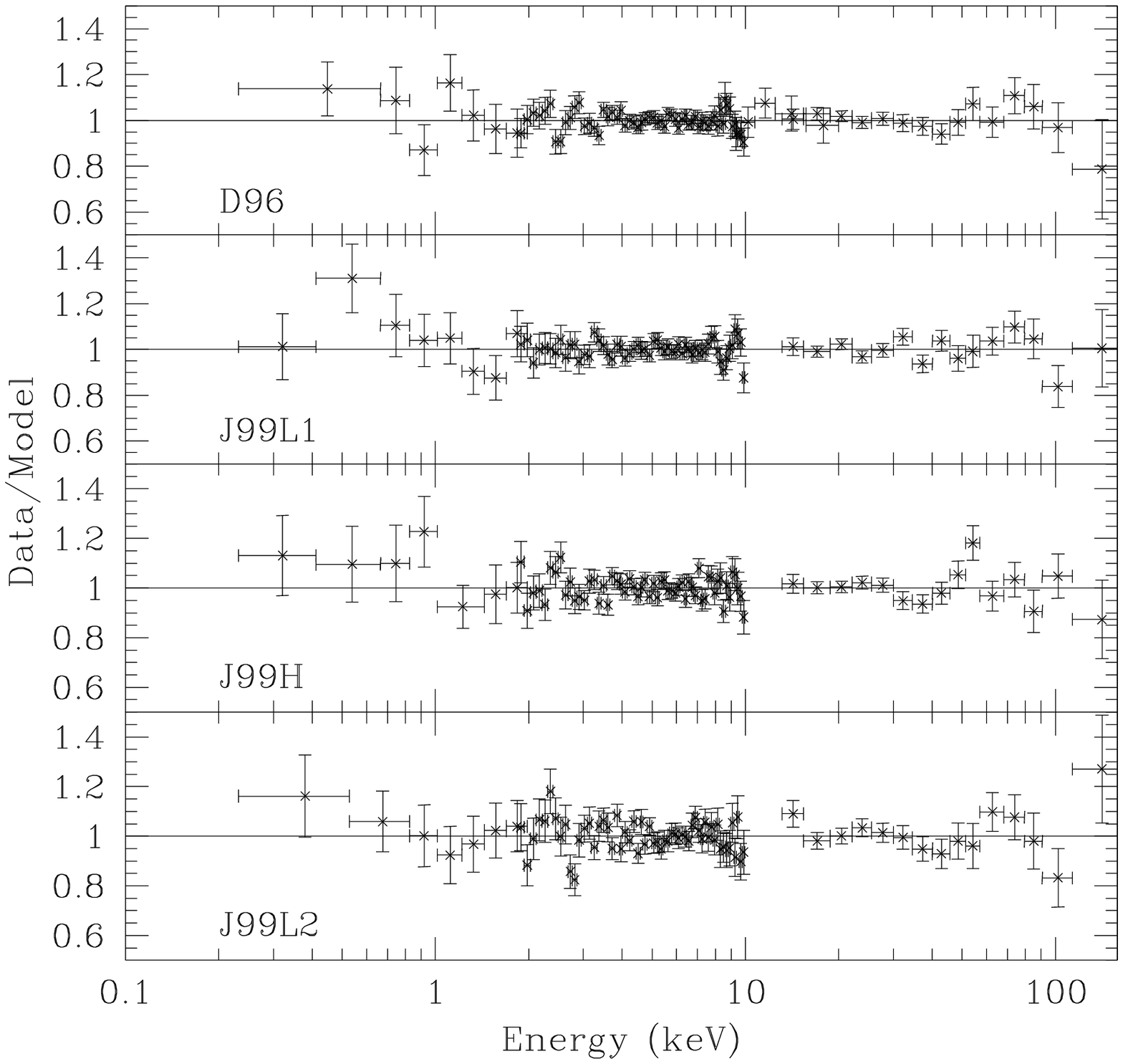}
 \caption{The Data/model ratios for the baseline model. }
\label{datamodel}
\end{figure*}

The BLM model is shown in Figure \ref{bbspec}. In the upper panel in
Figure the spectrum with model
of J96L is plotted with all the different spectral components while in
the lower panel two spectral states with a flux variation of a factor
of $\sim$ 3 (see Table \ref{bbfit}) are shown: December 2000 and December 2001.
We applied this template to all the \sax spectra in the different
flux level states (see Table \ref{log} and light curves in Figure
\ref{lc}). The best fit
parameters and the reduced $\chi^2$ values are shown in Table \ref{bbfit},
while the data/model ratio for each spectrum is plotted in Figure
\ref{datamodel}.
In the case of July 1996 high state (when LECS data are not available)
the parameters of the soft
components (thermal emission and scattering component) are frozen to
those derived for the spectrum in July 1996 low state.
%
\begin{sidewaystable*}
\centering
\caption[] {Broad band (1.8-200 keV) fits with our BLM. $E_{\rm C}$ is the
  e-folding energy of the intrinsic continuum, $A_{\rm IC}$ the
  normalizzation of the intrinsic continuum at 1 keV, $A_{\rm refl}$ is the
  strength of the Compton reflection hump obtained as
  $\Omega/2\pi \times A_{\rm IC}$, $I_{\rm Fe}$ is the flux of the
  iron line at 6.4 keV, $N_{\rm Hcov}$ and $N_{\rm Hwarm}$ are the column
  densities of the cold partially covered and warm gas respectively,
  $f_{\rm cov}$ is the covered fraction of the
  cold gas, $\xi$ is the ionization parameter of the warm gas,
  $EM_{\rm thermal}$ is the emission measure of the thermal gas - difined by $\int n_en_H
  dV$, where $n_{\rm e}$ is the electron density, and
  $n_{\rm H}$ is the hydrogen density - 
  assuming the distance to the source of 13 Mpc, and $A_{\rm
  scatt}$ is the 1 keV flux of soft scattered component.}
\begin{tabular}{l l l l l l l l l l l l l l l l }
\noalign {\hrule}
\noalign {\medskip}
 Period & $\Gamma$ & $^1E_c$ & $^2A_{\rm IC}$ & $^2A_{\rm refl}$ &
 $^3I_{\rm Fe}$
 & $(Z/Z_\odot)_{\rm Fe}$ & $^4N_{\rm Hcov}$ & $f_{\rm cov}$ &
 $^4N_{\rm Hwarm}$ &
 $^5\xi$ & $^7EM_{\rm therm}$ & $^8A_{\rm scatt}$ & $^6F_{2-10 {\rm keV}}^{unabs}$
 & $^6F_{0.1-200 \rm keV}^{unabs}$ & $\chi^2_{\nu}$\\ \hline

J96L & $1.58^{+0.37}_{-0.39}$ & $143^{+380}_{-84}$ &
$3.9^{+2.9}_{-2.0}$ & $2.4^{+4.2}_{-2.5}$ & $4.0^{+0.7}_{-0.5}$ &
$1.4^{+2.3}_{-0.5}$ & $22.1^{+5.1}_{-8.3}$ & $0.68^{+0.05}_{-0.08}$ &
$5.4^{+1.8}_{-1.8}$  & $8^*$ & $7.0^{+9.1}_{-7.0}$
&$2.1^{+0.7}_{-0.5}$ & 2.04 & 10.5 & 96.5/77 \\

J96H &$1.51^{+0.05}_{-0.22}$ & $81^{+40}_{-20}$ &$5.4^{+2.6}_{-1.4}$
&$1.0^{+0.6}_{-1.0}$ & $2.7^{+1.0}_{-1.0}$ & $1.8^{+0.7}_{-0.5}$ & $11.0^{+7.8}_{-3.7}$
&$0.49^{+0.25}_{-0.14}$ &$2.1^{+1.0}_{-2.0}$ & $16^*$ & & & 3.01 & 13
& 76.5/67\\

D96 &$1.63^{+0.22}_{-0.27}$ & $130^{+221}_{-58}$ & $4.1^{+2.1}_{-0.5}$ &
$2.1^{+2.3}_{-1.3}$ &$4.1^{+0.5}_{-0.5}$ &$1.5^{+0.3}_{-0.3}$ &
$22.0^{+12.2}_{-5.2}$ & $0.57^{+0.04}_{-0.17}$ &
 $9.4^{+2.8}_{-2.2}$ &$9.3^*$ &$8.6^{+7.4}_{-6.7}$ & $2.2^{+0.6}_{-0.4}$&
2.01 & 11 & 64.1/77 \\

J99L1 &$1.59^{+0.18}_{-0.17}$ & $125^{+129}_{-32}$ & $5.5^{+2.3}_{-1.7}$ &
$0.27^{+1.2}_{-0.27}$ & $4.2^{+0.8}_{-0.8}$ & $1.7^{+0.6}_{-0.4}$ &
$21.0^{+6.4}_{-5.3}$ & $0.64^{+0.07}_{-0.09}$ & $8.5^{+1.6}_{-2.0}$ &
$8.5^*$ & $12.4^{+7.3}_{-5.5}$ & $1.8^{+0.5}_{-0.4}$ & 2.72 & 13 & 61.7/72 \\

J99H & $1.74^{+0.22}_{-0.24}$ &$177^{+2473}_{-83}$ &
$8.1^{+4.6}_{-2.9}$ & $1.3^{+1.6}_{-1.3}$ & $2.8^{+0.9}_{-0.9}$ &
$1.6^{+1.1}_{-0.6}$ &$19.8^{+5.6}_{-6.4}$ &$0.58^{+0.07}_{-0.08}$
&$6.8^{+1.4}_{-1.4}$ & $2.9^*$ &$14.3^{+7.3}_{-5.5}$ &
$2.5^{+0.8}_{-0.7}$ & 3.38 & 16.5 & 85.7/72\\

J99L2 &$1.56^{+0.27}_{-0.21}$ &$130^{+496}_{-53}$ &
$4.6^{+2.2}_{-1.7}$ & $0.6^{+2.0}_{-0.6}$ &$4.2^{+1.0}_{-0.9}$ &
$1.8^{+0.6}_{-0.9}$ &$12.3^{+5.4}_{-5.4}$ &$0.70^{+0.12}_{-0.32}$
& $6.4^{+3.7}_{-2.0}$ & $10.3^*$ &$3.5^{+6.7}_{-3.5}$ &
$2.0^{+0.7}_{-0.5}$ & 2.44 & 13 & 73.0/72 \\

D00 & $1.71^{+0.17}_{-0.23}$ & $>200$ & $2.4^{+2.8}_{-1.1}$
&$1.3^{+2.0}_{-1.3}$ & $2.1^{+1.4}_{-0.7}$ & $1.7^{+1.7}_{-1.1}$ &
$30^{+94}_{-24}$ & $0.34^{+0.21}_{-0.24}$ & $8.3^{+3.9}_{-4.2}$ &$7.8^*$
&$3.0^{+12.1}_{-3.0}$ & $1.5^{+0.6}_{-0.4}$ & 1.01 & 5.9 & 71.2/72\\

D01 & $1.52^{+0.02}_{-0.06}$ & $122^{+13}_{-11}$ &$4.7^{+0.3}_{-0.3}$
&$0.03^{+0.25}_{-0.03}$ & $2.8^{+0.4}_{-0.4}$ & $5.8^{+1.0}_{-2.1}$
& $3.5^{+1.1}_{-0.9}$ & $0.39^{+0.08}_{-0.07}$ & $0.9^{+0.2}_{-0.3}$ &$0.7^*$
&$4.2^{+6.5}_{-4.2}$ & $2.5^{+0.8}_{-0.7}$& 2.66 & 14 &71.4/72 \\

\noalign {\hrule}
\noalign{\medskip}
\noalign{\noindent
Note: $^{(1)}$ e-folding energy in keV;
$^{(2)}$ in $10^{-2}$  keV$^{-1}$ cm$^{-2}$ s$^{-1}$;
$^{(3)}$ in $10^{-4}$  cm$^{-2}$ s$^{-1}$ at the line;
$^{(4)}$ in $10^{22}$ cm$^{-2}$;
$^{(5)}$ in {\rm erg cm s$^{-1}$};
$^{(6)}$ in $10^{-10}$ $\flux$;
$^{(7)}$ in $10^{62}$ cm$^{-3}$ ;
$^{(8)}$ in $10^{-3}$   keV$^{-1}$ cm$^{-2}$ s$^{-1}$;
$^*$ These parameters are frozen to their values of best fit computing the errors. }

\end{tabular}
\label{bbfit}
\end{sidewaystable*}

The broad-band baseline model described above provides a good fit
to all the spectra of \41 (Table \ref{bbfit}). In the following sections we
discuss the various spectral components and their variability.

\subsection{The origin of the spectral variability in the 2--10 keV
range and the intrinsic power law}
\label{spectral variability}

The source has always shown - in all the observations performed so
far and including ours - a systematic trend: the spectrum gets
softer when the flux increases. The origin of this behaviour is
one of the key issue in this source.

The broad band fit (presented in  Table \ref{bbfit}) shows that
the only spectral parameters showing significant changes, which
can be related to the 2--10 keV spectral variability, are those  of
the absorbing media. Indeed, a decrement of the effective
absorption (i.e. of $N_{\rm H}$ and/or $f_{\rm cov}$) results in a softer
spectrum, as appears when comparing the spectral ratios presented
in Section \ref{specratios} with the best 
fit parameters in Table \ref{bbfit}.  For example,
the softening observed from J96L to J96H (Figure \ref{ratios}) is
reproduced by decreasing the column densities of the two absorbers
and the covering fraction (Table \ref{bbfit}). On the contrary the
intrinsic photon index does not show significant evidence of
variation around the average value of $\Gamma\approx1.6$. A
variation $\Delta\Gamma\approx 0.2$ would still be
 consistent with the errors, but it could not itself account 
for the 
large spectral changes observed in this
range.  While most of the spectral variability
observed in the 2--10 keV range is due to variation of the
absorbers, a minor effect of intrinsic variations can not be
excluded.
This is demostrate by fitting the spectra with the intrinsic photon
index given by the F$_{2-10 \rm keV}$ $vs$ $\Gamma$ relation found 
by Perola \etal (1986): 
$\Gamma=2.18+0.012 \times F_{2-10 \rm keV}$
with $F_{2-10 \rm keV}$ in $10^{-11}$ erg cm$^{-2}$ s$^{-1}$.
The results of  these fits are reported in 
Table \ref{bbfit gamma fix} and show that in all spectra a good
$\chi^2_\nu$ is obtained.

\begin{sidewaystable*}
\caption[] {As Table \ref{bbfit} with
  the intrinsic photon index $\Gamma$ frozen following the F$_{2-10
  \rm keV}$ vs 
  $\Gamma$ relation found by \cite {per86}.}
\begin{tabular}{l l l l l l l l l l l l l l l}
\noalign {\hrule}
\noalign {\medskip}
 Period & $\Gamma$ & $^1E_c$ & $^2A_{\rm IC}$ & $^2A_{\rm refl}$ &
 $^3I_{\rm Fe}$
 & $(Z/Z_\odot)_{\rm Fe}$ & $^4N_{\rm Hcov}$ & $f_{\rm cov}$ &
 $^4N_{\rm Hwarm}$ & $^5\xi$ & $^7EM_{\rm therm}$ & $^8A_{\rm scatt}$
 & $^6F_{2-10 \rm keV}^{unabs}$
& $\chi^2_\nu$\\ \hline

J96L & $1.42$ & $68\pm 5$ & $3.0^{+0.2}_{-0.2}$ 
& $1.1^{+0.4}_{-0.4}$ & $3.9^{+0.5}_{-0.5}$ &
$2.3^{+0.4}_{-0.3}$ & $20.1^{+4.3}_{-2.7}$ & $0.66^{+0.04}_{-0.06}$ &
$5.1^{+1.7}_{-1.1}$  & $10^*$ & $8.1^{+7.6}_{-5.7}$ &$2.2^{+0.6}_{-0.5}$ 
& 2.04 & 96.4/78 \\

J96H &$1.54$ & $86^{+11}_{-11}$ &$5.2^{+0.7}_{-0.5}$
&$1.6^{+0.9}_{-0.9}$ & $2.9^{+0.9}_{-0.9}$ & $2.9^{+1.4}_{-1.1}$ & 
$10.4^{+6.2}_{-7.3}$ &$0.50^{+0.18}_{-0.09}$ &$2.6^{+0.8}_{-1.1}$ & 
$16^*$ & & & 3.01 & 73.2/68\\

D96 &$1.42$ & $79^{+10}_{-10}$ & $2.8^{+0.3}_{-0.2}$ & $1.0^{+0.4}_{-0.4}$ 
&$4.1^{+0.5}_{-0.5}$ &$2.0^{+0.4}_{-0.4}$ & $18.1^{+10.1}_{-4.5}$ & 
$0.54^{+0.10}_{-0.13}$ & $8.3^{+2.2}_{-1.8}$ &$8.3^*$ &
$10.0^{+5.2}_{-4.7}$ & $2.0^{+0.3}_{-0.3}$ & 2.01 & 65.6/78 \\

J99L1 &$1.51$ & $100^{+10}_{-10}$ & $4.6^{+0.4}_{-0.2}$ &
$>0.5$ & $4.2^{+0.7}_{-0.8}$ & $1.8^{+0.4}_{-0.3}$ &
$19.6^{+5.5}_{-2.4}$ & $0.63^{+0.07}_{-0.08}$ & $7.9^{+0.9}_{-2.5}$ &
$8.7^*$ & $13.0^{+6.1}_{-7.0}$ & $1.8^{+0.4}_{-0.4}$ & 2.72 & 61.9/73 \\

J99H & $1.59$ &$122^{+15}_{-13}$ &
$6.5^{+0.5}_{-0.4}$ & $0.3^{+0.6}_{-0.3}$ & $2.9^{+0.9}_{-0.9}$ &
$2.1^{+0.5}_{-0.4}$ &$17.3^{+4.9}_{-3.3}$ &$0.56^{+0.06}_{-0.07}$
&$6.3^{+1.1}_{-0.9}$ & $2.7^*$ &$14.4^{+7.5}_{-5.4}$ &
$1.5^{+0.5}_{-0.4}$ & 3.38 & 86.2/73\\

J99L2 &$1.47$ &$100^{+14}_{-11}$ &
$3.9^{+0.3}_{-0.3}$ & $0.2^{+0.6}_{-0.2}$ &$4.2^{+0.9}_{-0.9}$ &
$2.1^{+0.7}_{-0.6}$ &$10.9^{+6.2}_{-2.6}$ &$0.67^{+0.15}_{-0.28}$
& $6.5^{+3.8}_{-2.0}$ & $10.9^*$ &$5.1^{+6.5}_{-5.1}$ &
$2.0^{+0.7}_{-0.5}$ & 2.44 & 73.1/73 \\

D00 & $1.30$ & $110^{+25}_{-25}$ & $1.0^{+0.1}_{-0.1}$ & $<0.1$ & 
$1.9^{+0.5}_{-0.5}$ & $8^{+11}_{-5}$ &
$2.1^{+1.6}_{-1.3}$ & $1.0^{+0.0}_{-0.6}$ & $1.6^{+3.3}_{-1.4}$ &$10^*$
&$2.3^{+4.4}_{-1.6}$ & $1.3^{+0.3}_{-0.3}$ & 1.01 & 74.8/73\\

D01 & $1.50$ & $122^{+7}_{-5}$ &$4.7^{+0.1}_{-0.1}$
&$<0.2$ & $2.8^{+0.4}_{-0.4}$ & $7.1^{+2.2}_{-2.1}$
& $3.1^{+0.9}_{-0.6}$ & $0.41^{+0.09}_{-0.07}$ & $0.8^{+0.2}_{-0.2}$ &$0.3^*$
&$2.8^{+2.3}_{-1.2}$ & $2.3^{+0.5}_{-0.9}$& 2.66 & 70.3/73 \\

\noalign {\hrule}
\noalign{\medskip}
\noalign{\noindent
Note: $^{(1)}$ e-folding energy in keV;
$^{(2)}$ in $10^{-2}$ keV$^{-1}$ cm$^{-2}$ s$^{-1}$;
$^{(3)}$ in $10^{-4}$ cm$^{-2}$ s$^{-1}$ at the line;
$^{(4)}$ in $10^{22}$ cm$^{-2}$;
$^{(5)}$ in {\rm erg cm s$^{-1}$};
$^{(6)}$ in $10^{-10}$ $\flux$;
$^{(7)}$ in $10^{62}$ cm$^{-3}$; 
$^{(8)}$ in $10^{-3}$ keV$^{-1}$ cm$^{-2}$ s$^{-1}$;$^*$
These parameters are frozen to their values of best fit computing the errors. }
\end{tabular}
\label{bbfit gamma fix}
\end{sidewaystable*}

\subsection{The complex absorber}
\label{sect:absdata}

With the exception of the 2001 observation, the cold absorber is characterized
by $N_{\rm Hcov}=(1-3)\times 10^{23}$ \cmM2 and $f_{\rm cov}=0.3-0.7$,  similar
to that found by Weaver \etal (1994b), who adopted the same model of this
analysis on \asca data. The model employed for the warm
gas with $\xi\sim$ 1--16 \rerg \, \rcm \,\rm s$^{-1}$ indicates an
iron edge at energy around 7.3--7.6 keV, consistently with
other measurements (\cite{yaq93}; \cite{wa}; \cite{wb}; \cite{schurch03}).
%
%
%
With the exception of D01, the iron abundance is $\approx 1.5$ times 
greater than the solar value. We will show later how D01
is peculiar in respect to several points.

With regard to the  variability of absorbers, we note the
following. 
Due to the fit with such a complex model, the errors on the
absorbers parameters are wide. However we note that the column
density of the warm absorber appears to show the most significant
variations in respect to those of the cold absorber.
We have therefore chosen to
fix the column density of the cold absorber to the average value
$N_{\rm Hcov}=15\times10^{22}$ \cmM2. This is consistent, within the
errors, with all the best fit values except D01. In addition we have fixed the spectral index, 
as from the above discussion, to the value 1.6. The result and all
the best fit parameters are reported in Table \ref{fit gamma
cold fix}. 
The $\chi^2$ is still acceptable.
The column density of the warm absorber
changes over a time scale of month-years, while it is consistent
with being constant over a day time scale.
Specifically, in July 1996, $N_{\rm Hwarm}\approx 3.5\times10^{22}$ \cmM2,
in  D96 to D00 $N_{\rm Hwarm}\approx 7\times10^{22}$ \cmM2.
On the contrary the covering fraction of the cold absorber shows
variability in a day time scale. For example changing from 0.73
to 0.48 in the low and high state of July 1996. 

The D01 spectrum appears to be rather peculiar with respect to the
other states. 
We already noted that in this state the iron abundance is about 5 times that
solar. In D01 the softening is reproduced 
with a substantial decrease of the thickness of both absorbers, 
with $N_{\rm Hcov}=3.5\times10^{22}$ \cmM2,
$N_{\rm Hwarm}=0.9\times10^{22}$ \cmM2. The covering fraction was also very low,
$f_{\rm cov}\approx 0.3$. 
Basically, in the state observed by \sax in December 2001, most of the bare nuclear
continuum came into view (Figure \ref{bbspec}, lower panel).
The D01 spectrum is peculiar also for the presence of an absorption feature at
$\sim$ 9 keV. This state of the source is described and
discussed in Piro \etal (2005).

\begin{table*}
\begin{center}
\caption[] {As Table \ref{bbfit}, without less important component, with
$\Gamma$=1.6 and the cold column density $N_{\rm Hcov}=15\times10^{22} \rcm^{-2}$.}
\begin{tabular}{l l l l l l l l l}
\noalign {\hrule}
\noalign {\medskip}
 Period & $^1E_c$ & $^2A_{\rm IC}$ & $^2A_{\rm refl}$ & $^3I_{\rm Fe}$
 & $f_{\rm cov}$ & $^4N_{\rm Hwarm}$ &
 *$^5\xi$ & $\chi^2_\nu$\\ \hline

J96L & $106^{+10}_{-10}$ &$3.3^{+0.2}_{-0.1}$ & $3.3^{+0.6}_{-0.3}$ &
$4.2^{+0.4}_{-0.4}$ & $0.73^{+0.03}_{-0.04}$ & $4.4^{+1.4}_{-0.9}$ &
$8^*$ & 104/79 \\

J96H & $98^{+14}_{-11}$ &$6.6^{+0.7}_{-0.8}$ &$1.2^{+0.6}_{-0.6}$
&$3.4^{+1.1}_{-0.5}$ & $0.48^{+0.05}_{-0.09}$ &$3.1^{+1.9}_{-0.8}$
&$16^*$ & 73/69 \\

D96 & $124^{+15}_{-15}$ &$3.5^{+0.2}_{-0.2}$ & $2.2^{+0.5}_{-0.3}$ &
$4.1^{+0.5}_{-0.4}$ & $0.60^{+0.10}_{-0.08}$ & $8.0^{+2.1}_{-1.5}$ &
$9.9^*$ & 76/79\\

J99L1 & $135^{+18}_{-15}$ & $5.0^{+0.3}_{-0.3}$ & $1.0^{+0.4}_{-0.4}$
& $4.2^{+0.7}_{-0.7}$ & $0.73^{+0.04}_{-0.06}$ & $6.9^{+1.6}_{-1.2}$
&$8.6^*$ & 73/74 \\

J99H & $127^{+16}_{-14}$ & $6.0^{+1.0}_{-0.7}$ & $0.6^{+.5}_{-0.5}$ &
$2.9^{+0.9}_{-0.9}$ & $0.60^{+0.04}_{-0.05}$ & $6.0^{+1.0}_{-0.7}$ &
$3.2^*$ & 87/74 \\

J99L2 & $144^{+30}_{-20}$ & $5.1^{+0.4}_{-0.4}$ & $0.8^{+0.6}_{-0.5}$
& $4.2^{+0.9}_{-0.9}$ & $0.73^{+0.07}_{-0.16}$ & $7.4^{+3.0}_{-1.5}$
&$9.7^*$ &74/74 \\

D00 & $422^{+3965}_{-225}$ & $1.8^{+0.1}_{-0.3}$ & $<0.5$ &
$2.0^{+0.5}_{-0.6}$ & $0.34^{+0.14}_{-0.28}$ & $6.7^{+3.1}_{-1.7}$
&$14^*$ &71/74 \\

D01 &$162^{+12}_{-11}$ & $6.0^{+0.2}_{-0.2}$ & $0.3^{+0.3}_{-0.2}$
& $2.6^{+0.5}_{-0.4}$ & $0.28^{+0.02}_{-0.03}$ &$1.9^{+0.2}_{-0.1}$
&$2.8^*$ & 107/74 \\
\noalign {\hrule}
\noalign{\medskip}
\noalign{\noindent
Note: $^{(1)}$ keV;
$^{(2)}$ in $10^{-2}$  keV$^{-1}$ cm$^{-2}$ s$^{-1}$;
$^{(3)}$ in $10^{-4}$  cm$^{-2}$ s$^{-1}$ at the line;
$^{(4)}$ in $10^{22}$ cm$^{-2}$}
\end{tabular}
\label{fit gamma cold fix}
\end{center}
\end{table*}

\subsection{The reflection component}
\label{reflection}

The Compton reflection component is significantly detected in 1996
(both in July and December, with a mean value of the {\it absolute}
normalization $A_{\rm refl}= (2.2\pm 0.3) \times 10^{-2}\ \rm{ph\
cm^{-2}\ s^{-1}\ keV^{-1}}$), it is only marginally detected in
1999 ($A_{\rm refl}= (0.8\pm 0.3)\times 10^{-2}\ \rm{ph\ cm^{-2}\
s^{-1}\ keV^{-1}} $), while it is consistent with zero in 2000 and
2001 (see Table \ref{fit gamma cold fix}). This result remains true 
even when the photon index is not fixed (see Table 
\ref{bbfit} and Table \ref{bbfit gamma fix}). Thus we find that the
reflection component is variable on time scales of years, while
there is not strong evidence of variability on the (intra) day time
scales that characterize the direct continuum. We can actually exclude
variations correlated with those of the intrinsic continuum on such
short time scales.

\subsection{High-energy spectral variability}

In  Section \ref{sect:absdata} we have concluded that the spectral
variability observed in the 2--10 keV range is dominated by
variations of the absorber.
At higher energies, although the effect of absorption is negligible,
the assessment  is  also not straightforward due to the presence
of the reflection component.  For example, a {\it constant}
absolute reflection component, added to a steep (constant slope)
variable continuum (\cite{sw02}), may also qualitatively mimic the 
$\Gamma$ $vs$ $F_{2-10 \rm keV}$ correlation. It is then particulary relevant to
discuss the high energy spectra in 2000 and 2001 when the
reflection intensity is consistent with zero and then any
intrinsic spectral variation can be observed without ambiguity. 
Between 10 keV and 200 keV
the spectral ratio D01/D00 is not consistent with a constant
($\chi^2_\nu=1.8$ per 15 degree of freedom), but shows a decreasing trend that can be
reproduced with a change of the intrinsic slope between the two states 
of about $0.17\pm 0.06$ (with the assumption of a constant high energy
cut-off). 
This result is that expected on the basis of the correlation $\Gamma$ $vs$
$F_{2-10 \rm keV}$, where $F_{2-10 \rm keV}$ is changing by a factor of three between D00 and
D01. 

\begin{figure*}[t]
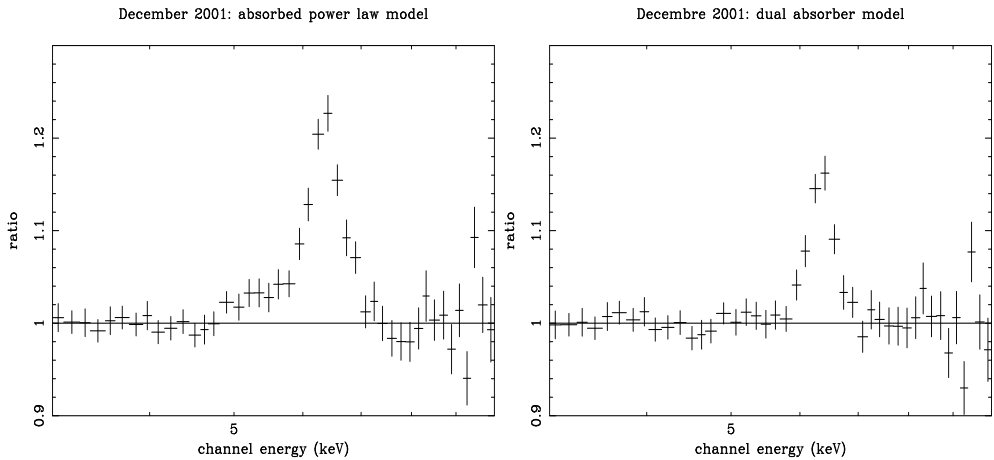

\centering
\includegraphics[height=6.5cm,width=6.cm,angle=-90]{iron_abspo_me01.ps}
\includegraphics[height=6.5cm,width=6.cm,angle=-90]{iron_d01_pc.ps}
 \caption{In the left panel we show the iron line profile when the continuum
 is reproduced with a simple absorbed power law in 3-10 keV. Clear
 positive residuals are evident between 5-6 keV. When the continuum
 is correctly modeled with a dual absorber model (our BLM), the
 residuals disappear, on the right panel, and a narrow gaussian
 profile becomes a good description to the iron line. The broadening
of  the profile in right panel is due to the MECS resolution (8
 per cent at 6 keV).}
\label{iron profile}
\end{figure*}

\subsection{The iron line}
\label{sec:linedata}

We recall that, in our modelling, we have adopted a narrow
gaussian profile for the line without any further component,
following the results of \chandra and \xmm
(\cite{chandra}; \cite{schurch03}). Because the issue of the existence and
origin of broad components in Seyfert galaxies is still under
debate, it is relevant to underline how this is dependent upon a
good modelling of the continuum, available thanks to the \sax broad-band data.
In Figure \ref{iron profile} we show 
the residuals when the continuum is modelled by a simple uniform absorbed
power law and the line shows evidence of a ``red wing''. However,
when the continuum is properly described with a dual absorber,
the ``red wing'' -- which is in fact due to the covered component --
disappears.

\begin{figure}
\centering
\includegraphics[height=6.5cm,width=.5\textwidth]{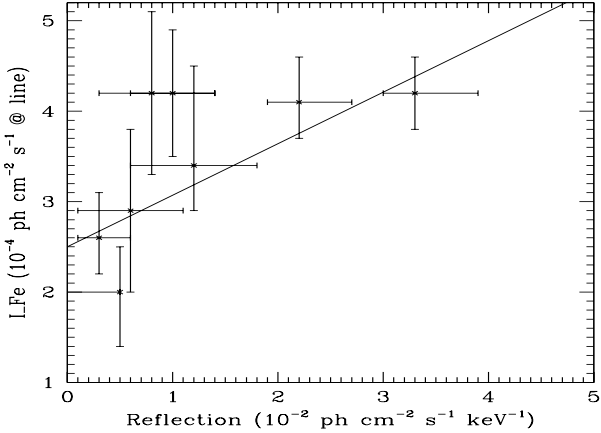}
\caption{The flux of  the Iron line is plotted as a function of the
Compton reflection component. We plot also the result of a linear fit} 
\label{fecorr}
\end{figure}

The flux of the iron line in our spectra shows evidence of
variability ($\chi^2/dof=19.6/7$ for a constant, chance probability P=0.0065), but these
variations are not correlated with the intensity of the intrinsic
continuum (correlation coefficient r=0.099). Interestingly the
flux of the line is lower in D00 and D01 when the reflection 
component disappears. To explore this
issue we plot in Figure \ref{fecorr} the line $vs$ reflection
intensity. A linear fit, plotted in the figure, 
is acceptable ($\chi^2/dof=7.8/6$, P=0.253) with
$I=(2.5\pm0.3)\times 10^{-4}+(0.57\pm0.17)\times 10^{-2}A_{\rm refl}$,
$I$ is in \pflux and $A_{\rm refl}$ in $\rm{ph\ cm^{-2}\ s^{-1}\
keV^{-1}}$. This fit shows that when the reflection component is
null, the line flux does not go to zero, identifying the
presence of a constant line component in addition to the one that
correlates with the reflection. While it is not statistically  
significant, it is nonetheless interesting to note that  for large
value of the reflection normalization, the line appears
to saturate. This point will be discussed later.

\subsection {The soft components}
\label{sec:softdata}

In the soft X-ray band the spectrum is well described by the
combination of a power law continuum and a thermal plasma with
$kT$=0.15 keV (see Table \ref{bbfit}). In Table \ref{tabsoft} we report
the flux of these two components in the 0.1--1 keV range.  
The total flux in this range is is dominated ($\approx$ 80 per cent) by
a constant power law with $F(0.1-1\ \rm keV)\approx
5\times10^{-12}\ \flux$, the remaining being attributed to the
thermal component. We have computed the expected flux of the
variable absorbed continuum in this energy range, finding that
it is at most 1.5 per cent of the total.
The only exception is the soft spectrum  D01 where there is a contribution of 
the intrinsic continuum in 0.1--1 keV due to the effect of low absorption
(see Table \ref{bbfit}).

\begin{table}
\begin{center}
\caption[] {Flux in 0.1--1 keV energy range.
We report the total flux and that of the thermal and power law component.
All the values are corrected for Galactic absorbtion.}
\begin{tabular}{l l l l }
\noalign {\hrule}
\noalign {\medskip}
 Period & $F^{tot}$ & $F^{therm}$ & $F^{scatt}$ \\ \hline

J96L & 6.6 & 1.4 & 5.1 \\


D96 & 6.6 & 1.2 & 5.4 \\

J99L1 & 6.7 & 2.3 & 4.4 \\

J99H & 6.8 & 3.0 & 3.8 \\

J99L2 & 5.4 & 0.6 & 4.8 \\

D00 & 4.7 & 0.7 & 4.0 \\

D01 & 8.1 & 1.6 & 5.5 \\

\noalign {\hrule} \noalign{\medskip} \noalign{\noindent Note: Flux
corrected for Galactic absorption, in $10^{-12}$ erg cm$^{-2}$
s$^{-1}$. The J96H state in not reported because the missing of soft
X-ray data from LECS (see Section \ref{observations}).}
\end{tabular}
\label{tabsoft}
\end{center}
\end{table}

\section{Discussion}
\label{discussion}

\subsection{The primary emission from the innermost region}

The intrinsic continuum of \41 is well described by a power law
with an exponential cut-off. The values of the high energy cut-off
we found in the different flux level states are all consistent
with the values observed in the Sy 1s by \sax (\cite {per02}).
This, together with the measure of the Compton reflection features
that perfectly match with the averaged values observed in Sy 1s
(\cite{nan94}), suggests that \41 is an intrinsically average
Seyfert 1 galaxy (\cite{zdz02}). The phenomenological model we
employed to fit the intrinsic continuum (i.e. power-law with high
energy cut-off), is a good description of a two-phase model
involving a hot corona emitting medium-hard X-rays by
Comptonization and a cold optically thick layer (the disk) that
provides the soft photons to Comptonization
(\cite{HM91}; \cite{haardt97}). 
While the dual absorber is the main driver of the observed variability
in 2--10 keV,
we find small amplitude of intrinsic variations above 10 keV,
that can be due to a variation of $\Delta\Gamma \approx
0.2$, according to the the correlation $\Gamma$ $vs$
$F_{2-10 \rm keV}$ (Perola \etal 1986). 
Assuming a distance of 13 Mpc for the source, and a black hole mass of
4$\times 10^7$ $\Ms$ we find that the ratio L$_{\rm {0.1-200
 keV}}^{\rm unabs}$/L$_{\rm E}$
between the unabsorbed luminosity and the Eddington luminosity is in
the range 0.002-0.006 in the lowest (D00) and highest (J99H) state
respectively. The accretion is taking place in a sub Eddington regime.

\subsection{The narrow iron line and the reprocessing region: NLR and molecular torus}

The energy of the narrow line as determined from the average of
all \sax observations is  $(6.40\pm0.06)$ keV, i.e. consistent
with fluorescence by cold iron. During our observations, the
line flux shows evidence of variations (see Table \ref{fit gamma cold fix}).
We find that the flux of the line is dependent upon the
normalization of the reflection component, and that it attains a
non-zero value ($I=(2.5\pm0.3)\times 10^{-4}$ \pflux) when the
reflection is absent (in Figure \ref{fecorr} we plot the flux of the
iron line as a function of the reflection amplitude). This suggests that
the line is made up by
two components. We first discuss the origin of the component at
zero reflection. Our result indicates that this component is
consistent with being constant, thus suggesting an extended origin.
In addition, it should be produced by an optically thin gas. The
flux we derive for this component is remarkably similar to
the spatially resolved line emission region observed by \chandra
 (\cite{chandra}, $I=(1.8\pm0.2)\times 10^{-4}$ \pflux). This region is
coincident with the NLR, also in agreement with our conclusion of
an optically thin gas.

The equivalent width of iron line in the case of an optically thin
medium illuminated by an isotropic continuum is given by (e.g
\cite{inoue}; \cite{piro_riga}):

\begin{equation}
$$EW\approx 100 \frac{N_{\rm H}}{10^{23} {\rm cm^{-2}}} (\frac{Z}{Z_\odot})_{Fe} \frac{\Delta\Omega}{4\pi}\ {\rm eV}$$
\end{equation}

where $\Delta\Omega/4\pi$ is the net solid angle covered by the
medium (equal to the angular extension of the  region
$\Delta\Omega^{*}$  multiplied by $f_{\rm cov}$). We can estimate the
column density of the NLR $N_{\rm H}=fRn$, by adopting the typical
values derived from optical observations of the size  $R\approx 1\
$kpc, density ($n\approx 200-1000\ \rcm^{-3}$) and filling factor
($f\approx 10^{-4}$; e.g. \cite{robinson}). By substituting these
values in eq.(1) we derive an $EW\sim 0.1-0.3$ eV that is  2 orders of magnitude
lower than observed (100-200 eV see Table \ref{bbfit}). One
possibility is that the ionizing flux is emitted anisotropically.
Penston \etal (1990) showed that the ionizing flux  needed to
produce the observed optical and UV line emission had to be about
10 times stronger than that observed. Explanations for this excess include
an anisotropic radiation field. But even in such a case,
the expected $EW$ would be lower  of the observed value
by a factor of 10. The presence of  a system of clouds with higher
density ($n\approx 10^6\ \rcm^{-3}$) can provide a more appealing
explanation. In this hypothesis the column density of NLR would be
$N_{\rm H}\sim 3\times 10^{23}$ \cmM2, then from eq. 1 we get an
equivalent width $EW\sim$ 200-300 eV, that is good agreement with the
values we found.
 Spectral diagnostics of narrow lines in a sample of
Seyfert nuclei that includes \41, reveals  the presence of low
density ($n\ltsima 10^4\ \rcm^{-3}$) {\it and} high ($n\gtsima 10^6\
\rcm^{-3}$) density clouds (\cite{stasinska}). Moreover, imaging
of the NLR of \41 with {\it HST} shows the presence of regions
whose density is likely to be enhanced by the shock compression of
the ejected radio material (\cite{winge}).
%

Let us  now discuss the line component that is correlated with the
reflection normalization, characterized by $I=(0.57\pm0.17)\times
10^{-2} A_{\rm refl}$ \pflux (see Figure \ref{fecorr}).
The line produced by reprocessing by a cold thick medium has an
intensity $I\approx 8\times 10^{-3} G Y[(Z/Z_\odot)_{\rm Fe}] A_{\rm refl}$ \pflux where
$G$ is a function depending on the geometry, and is equal to
$\approx 1$ (substantially independent on the inclination angle)
for simple configurations like a planar disk (\cite{mpp})
$Y[(Z/Z_\odot)_{\rm Fe}]$
describes the non-linear dependence of the line $EW$ upon the iron
abundance. For  $(Z/Z_\odot)_{\rm Fe}=2$, the case of \41, it is $Y=1.6$
(\cite{piro_riga}), and the line flux should be $I\approx
1.2\times 10^{-2} G A_{\rm refl}$ \pflux , remarkably similar to the
observed one for $G\approx 0.5$.
Values of $G$ less than unity are expected in a geometrically and
optically thick torus, where the reprocessing  takes place mostly
in the inner wall, i.e. in a cylindrical or conical geometry. This
is due to the ``secondary illumination'' effect, produced by
photons emerging from the wall at angles that intercepts the wall
again. This effect enhances the efficiency of reprocessing and
does it more for the photons of the reprocessed continuum than
for those in the line, because a 6.4 keV photon has a high
probability to be absorbed. For example, if the disk is in a
conical configuration with an opening angle of $45\deg$,
$G\approx0.5$ (\cite{mpp}).

\begin{figure}[]
\vspace{-0.5cm}
\centering
\includegraphics[height=6.5cm,width=0.5\textwidth]{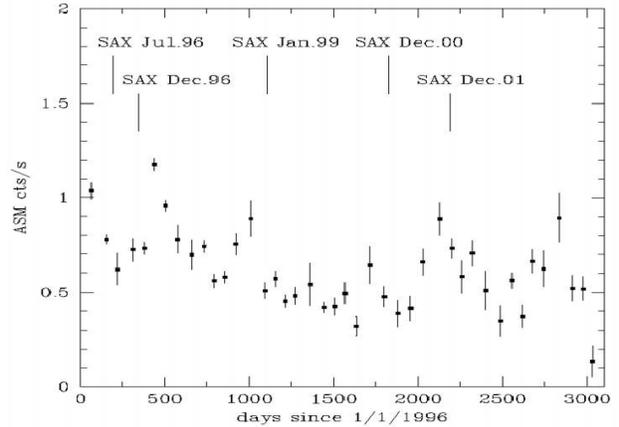}
\caption{Light curve in 1.5--10 keV of \41 from ASM aboard \rxte. Each bin is the
flux averaged on 60 days.} \label{fig:asm}
\end{figure}

We now discuss the origin of the variability of the reflection
component and the associated iron line. This could be due either
to a variation of the illuminating continuum or to a change of the
geometry of the reprocessing region. We recall that the intensity
of the reflection decreases on a time scale of years by a factor
of two or more from 1996 to 2000-2001 (see Table \ref{fit gamma
cold fix}). The size of the reprocessing region is therefore to be
of the order of light-years.  The intensity of the reflection
component thus traces the level of the illuminating continuum
averaged and delayed over a time scale of $\approx$ year. We have
therefore looked into the historical light curve of \41, to search
for evidence of long time scale variability. In Figure
\ref{fig:asm} we report the light curve as observed by the ASM
aboard of {\it RXTE}. Each point is the average flux over 2 months. The
source shows a systematic decrease in its intensity by almost a
factor of two from 1996 to 2000, consistent with the observed
reflection variability.

The most obvious candidate for the reprocessing region is a thick
torus whose inner walls are at $\approx$ light years away from the
central source. Taking then into account the line and reprocessing
intensities we estimate\footnote{A precise determination of the
geometry of the reprocessor goes however beyond the scope of this
paper. It would require detailed computations, that should take
into account the non-linear effect of secondary illumination, the
solid angle subtended by the inner surface of the torus towards
the X-ray source and their exposed area towards the observer.}
that the distance of the inner region of the torus should be
$\approx 3-4$ times its height. The ionizing continuum would
therefore emerge from the torus within a cone of opening angle of
$\approx 70 \deg$, in agreement with the estimation based on the
extended Narrow Line Region made by (\cite{pedlar}; \cite{robinson}).
Finally we note that the variability of the reprocessing component
could alternatively be explained by a change of the height of the
torus.

\subsection{The X-ray absorbers and the BLR}

\sax observations of \41 show the presence of a complex absorber
system. In particular the intrinsic continuum is absorbed by a
cold gas that partially obscure the central source and by a second
uniform warmer gas. Our analysis shows evidence of variations in
the status of the cold absorber on time scales of the order of the
day and of the warm gas on longer
time scale. Variations of the covered component can be produced if
the cold medium is made up of clouds passing in the line of sight
towards the central source (\cite{reichert}). If we identify these
clouds with the BLR, from the typical velocity of $\approx 10^4$
{\rm km s$^{-1}$}, we derive that the size of the central source
 $R\approx10^{14}$ {\rm cm}.
This would correspond, for $R\approx 10$ Schwarzschild radii, to a
black hole mass of $4\times 10^7\  \Ms$, consistent with
independent  estimations (\cite{clavel}; \cite{wandel}).  If the cloud
velocity is due to the orbital motion, we derive that the distance
of the clouds is a few light days, consistent with the size of the
inner BLR in NGC 4151 (\cite{clavel}).

In our fit we have associated the warm stage with a uniform screen
covering the source. In doing so we are guided by the arguments
put forward by Warwick \etal (1995) to solve the discrepancy
between the X-ray and UV measurements of absorption. Upper limits
on UV extinction (\cite{kriss}) correspond to a column of
$<8\times 10^{20}\ \rcm^{-2}$ assuming a standard gas-to-dust
ratio. This compares to X-ray columns of
$10^{22}-10^{23}\ \rcm^{-2}$.
In our case, the level of photoionization is
such that a substantial fraction of light elements as H, He and C
are ionized (e.g. \cite{krolik}), and therefore their absorption
which is predominant in the UV range is reduced to a level
compatible with UV measurements. We therefore associate the
uniform absorber to the warm medium, while the patchy absorber is
in a cold (neutral) stage.

Let us now characterize this warm medium we associated to the UV
line absorbing system (\cite{kriss}). From the relationship
$\xi=L/nR^2\approx10$ {\rm erg cm s$^{-1}$}, with
$L_{\rm {0.1-200 \rm keV}}\approx 3\times 10^{43}\ ${\rm erg s$^{-1}$} and the lower limit on the
density $n\gtsima10^{9.5}\ \rcm^{-3}$ derived by UV observations,
we obtain that the size of this region should be $R\ltsima 
10^{17}\ $\rcm. This is consistent with the upper limit derived
from the X-ray variability ($R\ltsima 10^{18}\ $\rcm) and the
requirement that the medium be external to the BLR ($R\gtsima
10^{16}\ $\rcm). 
In conclusion,  our
data  are fully consistent with the scenario in which the ionized
X-ray absorber coincides with the system responsible of UV
absorption lines. This system is composed by many clumps of dense
material covering the broad line region, and produced by an
outflow of the  broad-line clouds (\cite{kriss}).

The substantial decrease of the warm and cold absorbing gas in D01
together with the evidence, in the same spectrum, of an absorption
feature at $\sim$ 8.5-9 keV suggests the presence of a multi-phase
ionized absorber. This topic is widely discussed in Piro \etal (2005).

\subsection{Low energy diffuse components: scattering and
thermal}

The low energy emission in \41 remains constant on time
scales of the order of years, indicating an extended emission.
We reproduced the soft X-ray spectrum with two components. One,
dominated by emission lines, which is well fitted by a thermal
plasma with $kT\approx 0.15\ $ keV with abundances comparable to
solar. This is in agreement with \chandra observations
(\cite{chandra}) that demonstrated that most of the soft X-ray
flux is extended and dominated by lines from a photoionized and
collisionally ionized plasma. The other component is a power law
with a normalization $\approx 3-10$ per cent (see Table 3) of that of the  hard
power law, which is likely to be produced by electron scattering
of the central continuum, as in the case of Seyfert 2 galaxies
(\cite{antonucci}; \cite{matt_s2}).
The fraction of this component in respect to the continuum 
is strongly dependent from the intrinsic flux. We expect the scatterd fraction
increase when the source flux decreases if the nuclear continuum is
scattered into our line of sight. Weaver et al. (1994a)
in \asca observation of \41 found a fraction scattered of 3--4 per cent for the high state and 5--6
per cent for the low state. 
In our \sax observations the upper value of the ratio A$_{\rm scatt}/A_{\rm IC}$
$\approx$ 10 per cent is obtained in the D00
spectrum when the continuum level was lower in respect to the other
states.

\section{Conclusions}
\label{conclusions}

In this paper we presented the broad-band (0.1--200 keV) \sax
spectrum of \41. The source was observed 5 times from 1996 to 2001
with durations ranging from a day to four days. We find that the
spectrum is complex characterized by several components. The
continuum emission is well reproduced by a cut-off power law.

At energies less than 5 keV the continuum is strongly obscured by
a dual absorber: a cold component, that we associated with the BLR
clouds orbiting around a central source with a black hole with a
mass of $4\times 10^7 \Ms$, characterized by
$N_{\rm H}\approx1.5 \times 10^{23}$ \cmM2, a covering fraction
$f_{\rm cov}$=0.3--0.7 and variable on time scales of days to years,
and a second warmer uniform screen (with $N_{\rm H}=(1-8)\times
10^{22}$ \cmM2, variable on time scales of months-years),
photoionized by the continuum emission and that we argue to be 
coincident with the UV absorbing region. 

The low energies spectrum is reproduced by a combination of a
two components, which are not absorbed by the cold and warm gas: a
thermal one and a power law with the same slope of the intrinsic continuum
but with the intensity which is a few per cent.

During our observations the source shows strong time variability
energy dependent in time scales ranging from hours to years. In
0.1--2 keV it did not show any variability, while in 2--10 keV
the variation was up to a factor of eight at 3 keV and three at
10 keV. In the PDS energy range (13--200 keV) the larger
amplitude of variation was of a factor of three. We found that in
2--10 keV most of the spectral variability can be attributed
to a variation of the absorbers. 
Instrinsic spectral changes can explain some of
the smaller amplitude spectral variability above 10 keV.

We detected a Compton reflection component at high energy and the 
iron line at 6.4 keV.
The intensity of the reflection component is significantly
different from zero in 1996 while it disappears in 1999 and later
(2000--2001). 
The size of this region, implied by the variability argument, as well
as its optical thickness argue
for an association with a geometrically and optically thick torus
similar to that of Seyfert 2 galaxies.
The study of the historical light curve of \41 suggests that the
reflection component traces the level of the illuminating
continuum averaged and delayed over a time scale of $\approx$
year.
We found that the iron line flux is well described by two
components: one is correlated with the reflection intensity and is associated with the torus, the
other is constant $I=(2.5\pm0.3)\times 10^{-4}$ \pflux (in very good
agreement with \chandra measurement). This component has to be produced in a
diffuse optically thin gas, as confirmed by \chandra observations
that have shown that about 70 per cent of the line is
extended over a size of $\approx 1$ kpc coincident with the
extended NLR.

\begin{acknowledgements}

The authors wish to thank the referee A. Zdziarski for constructive criticisms which have greatly improved the paper.
A.D.R.  acknowledge financial contributin from ASI-INAF I/023/05/0.                                                                              
\end{acknowledgements}


\begin{thebibliography}{}

\bibitem[Anders \& Grevesse 1989]{}
Anders, E., \& Grevesse, N. 1989, GeCoA, 53, 197

\bibitem[Antonucci 1993]{antonucci}
Antonucci, R.R. 1993, ARA\&A, 31, 473


\bibitem[Beckmann \etal 2005]{beckmann2005}
Beckmann, V., Shrader, C.R., Gehrels, N.,\etal 2005, ApJ, 634, 939

\bibitem[Boella \etal 1997]{boella}
Boella, G., Butler, R.C., Perola, G.C., Piro, L., Scarsi, L., Bleeker,
J.A.M. 1997, \aas, 122, 299


\bibitem[Clavel \etal 1987]{clavel}
Clavel, J., Altamore, A., Boksenberg, A., \etal 1987 \apj, 321, 251

\bibitem[De Rosa \etal 2002]{derosa02}
De Rosa, A., Piro, L., Fiore, F. 2002a, A\&A, 387, 838

\bibitem[De Rosa \etal 2004]{derosa04}
De Rosa, A., Piro, L., Matt, G., Perola, G.C. 2004, A\&A, 413, 895


\bibitem[Done \etal 1992]{done92}
Done, C., Mulchaey, J.S., Mushotzky, R.F., Arnaud, K.A. 1992, ApJ 395, 275

\bibitem[Done \etal 2000]{done00}
Done, C., Madejski, G.M., Zycki, P.T. 2000, ApJ, 536, 213


\bibitem[Fabian \etal 2000]{fabian2000}
Fabian, A.C., Iwasawa, K., Reynolds, C.S., Young, A.J. 2000, PASP, 112, 1145


\bibitem[Fiore, Guainazzi \& Grandi 1999]{fgg99}
Fiore, F., Guainazzi, M., \& Grandi, P. 1999, {\it Cookbook for
BeppoSAX NFI Spectral Analysis}






\bibitem[Haardt \& Maraschi 1991]{HM91}
Haardt, F., \& Maraschi, L. 1991, ApJ, 380, L51

\bibitem[Haardt, Maraschi \& Ghisellini 1997]{haardt97}
Haardt, F., Maraschi, L., \& Ghisellini, G. 1997, \apj, 476, 620


\bibitem[Kaspi \etal 2001]{kaspi01}
Kaspi, S., Brandt, W.N., Netzer, H., \etal 2001, ApJ, 554, 216

\bibitem[Holt \etal 1980]{hol80}
Holt, S.S., \etal 1980, \apj, 241, L13

\bibitem[Inoue 1989]{inoue}
Inoue, H. 1989, proc. of {\it 23rd ESLAB symposium in X-ray
astronomy}, Bologna, Sep.13-20 1989, p.783 (ESA SP-296)


\bibitem[Krolik \& Kallman 1984]{krolik}
Krolik, J.H., \& Kallman, T.R., 1984, \apj, 286, 366

\bibitem[Kriss \etal 1992]{kriss}
Kriss, G.A., Davidsen, A.F., Blair, William P.,\etal 1992, \apj, 392, 485.



\bibitem[Matt, Perola \& Piro 1991]{mpp}
Matt, G., Perola, G.C., \& Piro, L. 1991, \aa, 247, 25

\bibitem[Matt \etal 1997]{matt_s2}
Matt, G., Guainazzi, M., Frontera, F., \etal 1997, \aa, 325, L13

\bibitem[Matt \etal 2001]{matt01}
Matt, G., Guainazzi, M., Perola, G.C., et al. 2001, A\&A, 377, L31

\bibitem[Magdziarz \& Zdziarski 1995]{pexrav_ref}
Magdziarz, P., \& Zdziarski, A.A. 1995, MNRAS, 273, 837


\bibitem[Nandra \& Pounds 1994]{nan94}
Nandra, K., Pounds, K.A. 1994, MNRAS, 268, 405

\bibitem[Nandra \etal 2000]{nandra00}
Nandra, K., Le, T., George, I., \etal 2000, ApJ, 544, 734

\bibitem[Nicastro \etal 2000]{nicastro00}
Nicastro, F., Piro, L., De Rosa, A., \etal 2000, ApJ, 536, 718

\bibitem[Ogle \etal 2000]{chandra}
Ogle, P.M., Marshall, H.L., Lee, J.C., \& Canizares, C.R. 2000 
ApJ, 545, L81

\bibitem[Page \etal 2002]{page02}
Page, K., Pounds, K., Reeves, J., O'Brien, P.T. 2002, MNRAS, 330, L1


\bibitem[Pedlar \etal 1983]{pedlar}
Pedlar, A., Kukula, M.J., Longley, D.P.T., \etal 1993, \mnras, 263, 471

\bibitem[Penston \etal 1990]{penston}
Penston, M.V., Robinson, A., Alloin, D., \etal 1990, \aa, 236, 53

\bibitem[Perola \etal 1986]{per86}
Perola, G.C., Piro, L., Altamore, A., et al. 1986, ApJ, 306, 508

\bibitem[Perola \etal 2002]{per02}
Perola, G.C., Matt, G., Cappi, M., Fiore, F., Guainazzi, M., Maraschi, L.,
Petrucci, P.O., Piro, L. 2002, A\&A, 389, 802



\bibitem[Petrucci \etal 2000]{petrucci00}
Petrucci, P.O., Haardt, F., Maraschi, L., \etal 2000, ApJ, 540, 131


\bibitem[Piro 1993]{piro_riga}
Piro, L. 1993, proc of {\it UV and X-ray spectroscopy of
astrophysical and laboratory plasmas}, Berkeley Feb.3-5 1992, E.H.
Silver \& S. M. Kahn ed.s, p. 448 (Cambridge Univ. Press)

\bibitem[Piro, Scarsi \& Butler 1995]{psb}
Piro, L., Scarsi, L., \& Butler, R.C. 1995, Proc. of SPIE, 2517,
169




\bibitem[Piro \etal 2005]{piro2005}
Piro, L., De Rosa, A., Matt, G. and Perola, G.C. 2005, A\&AL, 441, L13

\bibitem[Pounds \etal 1986]{pou86}
Pounds, K.A., Warwick, R.S., Culhane,J.L., de Korte, P. 1986, MNRAS 218, 685

\bibitem[Pounds \etal 2001]{pounds01}
Pounds, K., Reeves, J., O'Brien, P., et al. 2001, ApJ, 559, 181


\bibitem[Reeves \etal 2001]{reeves01}
Reeves, J.N., Turner, M.J.L., Pounds,K.A., \etal 2001, 365, L134

\bibitem[Reichert, Mushotzky \& Holt 1986]{reichert}
Reichert, G.A., Mushotzky, R.F., \& Holt, S.S. 1986 , \apj, 303, 87

\bibitem[Robinson \etal 1994]{robinson}
Robinson, A., Vila-Vilaro, B., Axon, D.J., \etal 1994, \aa, 291, 351

\bibitem[Schurch \& Warwick 2002]{sw02}
Schurch, N.J., \& Warwick, R.S. 2002, MNRAS, 334, 811

\bibitem[Schurch \etal 2003]{schurch03}
Schurch, N.J., Warwick, R.S., Griffiths, R.E., Sembay, S. 2003, MNRAS, 345, 423

\bibitem[Stasi\'{n}ska 1984]{stasinska}
Stasi\'{n}ska, G. 1984, \aa,135, 341

\bibitem[Svensson 1994]{svensson}
Svensson, R. 1994, \apjs, 92, 585

\bibitem[Tanaka \etal 1995]{tanaka}
Tanaka, Y., Nandra, K., Fabian, A.C., \etal 1995, Nature, 375, 659

\bibitem[Ulrich 2000]{ulrich_rev}
Ulrich, M.-H., 2000, A\&A Rev, 10, 135

\bibitem[Vaughan \& Edelson 2001]{vaughan01}
Vaughan, S., \& Edelson, R. 2001, ApJ, 548, 694


\bibitem[Wandel, Peterson \& Malkan 1999]{wandel}
Wandel, A., Peterson, B.M., \& Malkan, M.A. 1999, \apj, 526, 579


\bibitem[Warwick, Done \& Smith 1995]{wds}
Warwick, R.S., Done, C., \& Smith, D.A. 1995, MNRAS , 275, 100

\bibitem[Weaver \etal 1994a]{wa}
Weaver, K.,  Mushotzky, R.F., Arnaud, K.A., \etal 1994a, \apj, 423, 621

\bibitem[Weaver \etal 1994b]{wb}
Weaver, K., Yaqoob, T.,0 Holt, S.S., \etal 1994b, \apj, 436, L27

\bibitem[Winge \etal 1997]{winge}
Winge, C., Axon, D.J., Macchetto, F.D., \& Capetti, A. 1997, \apj
 487, L121

\bibitem[Yaqoob \etal 1991]{yaq91}
 Yaqoob, T., \& Warwick, R.S. 1991, MNRAS, 248, 773

\bibitem[Yaqoob \etal 1993]{yaq93}
 Yaqoob, T., Warwick, R.S., Makino, F., \etal 1993, MNRAS, 262, 435


\bibitem[Yaqoob \etal 2001]{yaq2001}
Yaqoob, T., George, I.M., Nandra, K., \etal 2001, \apj, 546, 759

\bibitem[Zdziarski \etal 1996]{zdz96}
 Zdziarski, A.A., Johnson, W.N., \& Magdziarz, P. 1996,
MNRAS 283, 193


\bibitem[Zdziarski \etal 2002]{zdz02}
 Zdziarski, A.A., Leighly, K.M., Matsuoka, M., Cappi, M., Mihara, T. 2002,
 ApJ, 573, 505

\bibitem[Zdziarski \etal 2003]{zdz03}
Zdziarski, A.A., Lubi\'{n}ski, P., Gilfanov, M., Revnivtsev, M. 2003, MNRAS, 342, 355

\end{thebibliography}
\end{document}